 \pdfoutput=1

\documentclass[journal]{IEEEtran}

\usepackage{url}
\usepackage{subfigure}
\usepackage{color}
\usepackage{wireless-utils}
\usepackage{cite}
\usepackage[pdftex]{graphicx}

\usepackage{amssymb}
\usepackage[cmex10]{amsmath}

\usepackage{mdwtab}

\usepackage{tabularx}
\usepackage{color}
\usepackage{xspace}
\usepackage{graphicx}
\usepackage{grffile}
\usepackage{times}
\usepackage{marvosym}
\usepackage[ruled,linesnumbered]{algorithm2e}

\usepackage{paralist}

\newtheorem{proposition}{Proposition}

\newtheorem{theorem}{Theorem}

\renewcommand{\add}[1]{\textcolor{black}{#1}}

\newcommand{\change}[1]{\textcolor{black}{#1}}

\graphicspath{{Fig/}}

\begin{document}

\title{Offloading Cellular Traffic through Opportunistic Communications: Analysis and Optimization}

\author{
Vincenzo Sciancalepore, Domenico Giustiniano, Albert Banchs, Andreea Picu
}

\maketitle
\begin{abstract}
Offloading traffic through opportunistic communications has been recently proposed as a way to relieve the current overload of cellular networks. Opportunistic communication can occur when mobile device users are (temporarily) in each other's proximity, such that the devices can establish a local peer-to-peer connection (e.g., via Bluetooth). Since opportunistic communication is based on the spontaneous mobility of the participants, it is inherently unreliable. This poses a serious challenge to the design of any cellular offloading solutions, that must meet the applications' requirements. In this paper, we address this challenge from an \emph{optimization analysis} perspective, in contrast to the existing \emph{heuristic} solutions. We first model the dissemination of content (injected through the cellular interface) in an opportunistic network with heterogeneous node mobility. Then, based on this model, we derive the optimal content injection strategy, which minimizes the load of the cellular network while meeting the applications' constraints. Finally, we propose an adaptive algorithm based on control theory that implements this optimal strategy without requiring any data on the mobility patterns or the mobile nodes' contact rates. The proposed approach is extensively evaluated with both a heterogeneous mobility model as well as real-world contact traces, showing that it substantially outperforms previous approaches proposed in the literature.
\end{abstract}

\IEEEpeerreviewmaketitle

\section{Introduction}

Following the huge popularization of smartphones and the ensuing explosion of mobile data traffic~\cite{gartner_mobile_devices}, cellular networks are currently overloaded and this is foreseen to worsen in the near future~\cite{cisco_mobile_traffic}. A recent promising approach to alleviate this problem is to offload cellular traffic through opportunistic communications~\cite{Pelusi}. The key idea is to inject mobile application content to a small subset of the interested users through the cellular network and let these users opportunistically spread the content to others interested upon meeting them. By exploiting opportunistic communications \add{in this way}, such an approach has the potential to substantially relieve the load of the cellular infrastructure. Among other mobile applications, this can be used for news~\cite{Ioannidis09optimaland}, road traffic updates~\cite{Whitbeck2011}, social data~\cite{Han2010} or streaming content~\cite{Keller}. \change{Indeed, as shown by our performance evaluation results, the load of the cellular network can be reduced between $50\%$ and $95\%$, depending on the application.}


Opportunistic networking exploits the daily mobility of users, which enables intermittent \emph{contacts} whenever two mobile devices are in each other's proximity. These contacts are used to transport data through the opportunistic network, which may introduce substantial delays. However, the type of content concerned by cellular offloading may not always be entirely delay-tolerant. In many applications, it is indeed critical that the content reach all users before a given deadline, lest it lose its relevance or its usability. Therefore, the design of opportunistic-based cellular offloading techniques faces serious challenges from the intermittent availability of transmission opportunities and the high dynamics of the mobile contacts. In order to find the best trade-off between the \emph{load} of the cellular network and the \emph{delay} until the content reaches the interested users, any opportunistic-based offloading design must answer crucial questions such as, \emph{how many} copies of the content to inject, to \emph{which users} and \emph{when}.


While a number of techniques have been proposed in the literature to offload cellular traffic through opportunistic communications, 
previous approaches are either based on heuristics (and hence do not ensure that the load of the cellular network is minimized)~\cite{Han2010,Whitbeck2011,Keller} or fail to provide delay guarantees~\cite{Ioannidis09optimaland,Keller}.
In contrast to the above approaches, in this paper, we propose the HYPE (HYbrid oPportunistic and cEllular) technique, which \emph{minimizes} the load of the cellular network while meeting the constraint in terms of \emph{delay guarantees}. To our best knowledge, we are the first to provide such features. The key contributions of our work are as follows:
\begin{compactenum}
\item Building on the foundations of \emph{epidemic analysis}~\cite{Daley}, we propose a model to understand the fundamental trade-offs and evaluate the performance of a hybrid opportunistic and cellular communication approach. Our model reveals that content tends to disseminate faster through opportunistic contacts when a sufficient, but not excessive, number of nodes have already received the content; in contrast, dissemination is slower when either few users have the content or few users are missing it.
\item Based on our model, we derive the optimal strategy for injecting content through the cellular network. In line with our previous findings, this strategy uses the cellular network when low speed of opportunistic propagation is statistically expected, and lets the opportunistic network spread the content the rest of the time.
\item We design an adaptive algorithm, based on control theory, that implements the optimal strategy for injecting content through the cellular network. The key strengths of this algorithm over previous approaches are that it adapts to the current network conditions without monitoring the nodes' mobility and that it incurs very low signaling overhead and complexity. Both features are essential features for a practical implementation.
\end{compactenum}

The rest of the paper is structured as follows. After thoroughly reviewing related work in Section~\ref{sec:related_work}, \change{we outline the basic design guidelines of our approach and theoretically analyze its performance in Section~\ref{sec:arch}. Based on this analysis, in Section~\ref{sec:opt_str}, we then derive the optimal strategy and present our adaptive algorithm, which implements this optimal strategy.} The algorithm's performance is extensively evaluated in Section~\ref{sec-perfeval}, using mobility models as well as experimental contact traces. Finally, Section~\ref{sec:conclusion} closes the paper with some final remarks. 

\section{Related Work}\label{sec:related_work}

The problem of the unsustainable increase in cellular network traffic and how to offload some of it has become more and more popular. Two types of solutions can be distinguished, on the basis of the outlet chosen for part of the cellular traffic: \begin{inparaenum}[(i)]\item offloading through additional (new or existing) infrastructure, and \item offloading through ad hoc communication\end{inparaenum}. Our proposal, HYPE, falls into the second category.

In the first category, many solutions~\cite{lee2010mobile,Tsao} are aiming to exploit the relatively large number of existing WLAN access points, as well as cellular diversity. A different approach, based on new infrastructure, is introduced in~\cite{Malandrino2012a}, in the context of vehicular networks. In that paper, the authors advocate the deployment of fixed roadside infrastructure units and study the performance of the system in offloading traffic information from the cellular network.

In the second category, along with our study, an increasing body of work is investigating the use of infrastructure-free opportunistic networking as a complement for the cellular infrastructure. In particular, the studies in~\cite{Han2010,Ioannidis09optimaland,Li2011,Keller,Whitbeck2011} propose solutions based on this idea.

In~\cite{Ioannidis09optimaland}, the authors propose to push updates of dynamic content from the infrastructure to subscribers, which then disseminate the content epidemically. The distribution of content updates over a mobile social network is shown to be scalable, and different rate allocation schemes are investigated to maximize the data dissemination speed. A substantial difference between this work and HYPE is that~\cite{Ioannidis09optimaland} does not minimize the load incurred in the cellular network and does not provide any delay guarantees, which are central objectives in our approach. Moreover, the solution introduced in~\cite{Ioannidis09optimaland} results in higher resource consumption for the ``most central'' users (i.e., highest contact rates) and/or the ``most social'' users.

Han et al. investigate, in~\cite{Han2010}, which initial subset of users (who receive the content through the cellular) will lead to the greatest infection ratio. A heuristic algorithm is proposed, that uses the history of user mobility of the previous day to identify a target set of users for the cellular deliveries. HYPE differs significantly from this, in the following aspects: \begin{inparaenum}[(i)]\item the solution in~\cite{Han2010} is heuristic and thus does not guarantee optimal performance, \item it requires to know the mobility patterns of all users, which may not be realistic in most scenarios, and \item it only investigates \emph{which} users to choose, but not \emph{how many} of them\end{inparaenum}.

In \cite{Keller}, an architecture is implemented to stream video content to a group of smartphones users within proximity of each other, using both the cellular infrastructure and WLAN ad-hoc communication. The decision of who will download the content from the cellular network is based on the phones' download rates. In contrast to our work, the focus of~\cite{Keller} is on the implementation rather than the model and the algorithm. Indeed, the algorithm proposed is a simple heuristic, which does not guarantee optimal performance.

Another study where opportunistic networking is used to offload the mobile infrastructure is~\cite{Li2011}. Here, some chosen users, named ``helpers'', participate in the offloading, and incentives for these users are provided by using a micro-payment scheme. Alternatively, the operator can offer the participants a reduced cost for the service or better quality of service. Thus, the focus of~\cite{Li2011} is on incentives, which is out of the scope of our work.

Most similar to HYPE is the Push-and-Track solution, presented in~\cite{Whitbeck2011}. There, a subset of users initially receive content from a content provider and subsequently propagate it epidemically. Upon reception of the content, every node sends an acknowledgment to the provider, which may decide to re-inject extra copies to other users. Upon reaching the content deadline, the system enters into a ``panic zone'' and pushes the content to all nodes that have not yet received it. The most prominent difference between this approach and ours is that Push-and-Track relies on a heuristic to choose \emph{when} to feed more content copies into the opportunistic network, which does not guarantee that the load on the cellular network is minimized. In contrast, we build on analytical results to guarantee that performance is optimal. An additional drawback of Push-and-Track is that it incurs a very high signaling overhead, which compromises the scalability with the number of subscribed users. Results in Section~\ref{sec-perfeval} confirm that our theory-driven algorithm outperforms the heuristics proposed in~\cite{Whitbeck2011} both in terms of cellular load and of signaling overhead.

Finally, from a different perspective, HYPE is also related to content dissemination solutions in purely opportunistic networks~\cite{Helgason,Hu2010}. However, most of these studies focus on finding the best ways to collaborate or contribute to the dissemination, under various constraints (e.g., limited ``public'' buffer space). Evaluation is usually based on the delay incurred to obtain desired content or the equivalent metric of average content freshness over time. In contrast, our metric is the load incurred in the cellular network. However, when developing our initial model, we do use a similar modeling method as in purely opportunistic dissemination (e.g.,~\cite{Helgason,Sermpezis2012}).

\change{Like all the previous works on offloading cellular networks through opportunistic communications~\cite{Ioannidis09optimaland, Whitbeck2011, Han2010}, with our approach all the transmissions over the cellular network are unicast. There are several key reasons that limit the usage of multicast messages in a cellular network. First, multicast cannot be easily combined with opportunistic transmissions, as this would require that the Content Server is aware of the cell of each node and can dynamically select the subset of nodes at each cell that receives the multicast message, which is not possible with current cellular multicast approaches. Second, in urban scenarios users will likely be associated to different base stations (there are hundreds/thousands of them in the city, each covering some sector, and in dense urban areas femtocells have started to be deployed). Thus, there is a low probability that users subscribed to a specific content are associated to the same cells at the same time, and hence multicast may collapse to unicast. Finally, transmissions with multicast would occur at the lowest rate to preserve users in the edge of the cell, which degrades the resulting performance.}

\section{The HYPE approach}\label{sec:arch}

In this section, we present the basic design guidelines of the HYPE (HYbrid oPportunistic and cEllular) approach. HYPE is a hybrid cellular and opportunistic communications approach that delivers content to a set of users by \begin{inparaenum}[(i)]\item sending the content through the cellular network to an initial subset of the users (which we will call \emph{seed nodes}), and \item letting these initial users or seed nodes share the content opportunistically with the other nodes\end{inparaenum}. We aim at designing HYPE so as to combine the cellular and opportunistic communication paradigms in a way that retains the key strengths of each paradigm, while overcoming their drawbacks.

HYPE consists of two main building blocks: \begin{inparaenum}[(i)]\item the \emph{Content Server}, and \item the \emph{Mobile Applications}\end{inparaenum}. The Content Server runs inside the network infrastructure, while the Mobile Applications run in mobile devices that are equipped with cellular connectivity, as well as able to directly communicate with each other via short range connections (e.g., via WLAN or Bluetooth). The Content Server monitors the Mobile Applications and, based on the feedback received from them, delivers the content through the cellular network to a selected subset of Mobile Applications (the seed nodes). When two mobile devices are within transmission range of each other, the corresponding Mobile Applications opportunistically exchange the content by using local (short-range) communications.

\subsection{Objectives}\label{subsec:obj}
The fundamental challenge of the HYPE approach is the design of the algorithm that decides \emph{which} mobile devices and \emph{when} they should receive the content through the cellular network. The rest of this paper is devoted to the design of such an algorithm. The key objectives in the design are:
\begin{compactenum}[(i)]
\item\label{it:offload} \emph{Maximum Traffic Offload}: Our fundamental objective is to maximize the traffic offloaded and thus reduce the load of the cellular network as much as possible. This is beneficial both for the operators (who may otherwise need to upgrade their network, if the cellular infrastructure is not capable of coping with current demand), as well as for the users (who must pay for cellular usage, either directly or by seeing their data rate reduced).
\item\label{it:delay} \emph{Guaranteed delay}: Most types of content have an expiration time, arising either from the content's usefulness to the user (e.g.,~road traffic information), its validity after an update (e.g.,~daily news) or its play-out time (e.g.,~streaming). Therefore, a key requirement for our approach is that the content reaches all the interested users before its deadline.
\item\label{it:fair} \emph{Fairness among users}: In order to make sure that all users benefit from HYPE, it is important to guarantee a good level of fairness both in terms of cellular usage (for which users have to pay), as well as in terms of opportunistic communications (which may increase the energy consumption of the device).\footnote{Indeed, an important drawback of certain existing solutions is that they tend to over-exploit the users with high contact rates~\cite{Ioannidis09optimaland,Han2010}, thus discouraging the participation of such users.}
\item\label{it:signal} \emph{Reduced signaling overhead}: The signaling overhead between the Content Server and the Mobile Applications needs to be low. This is important for two reasons: first, to ensure that HYPE \emph{scales} with the number of mobile devices (otherwise the signaling traffic would overload the cellular network); second, to avoid using the cellular interface for small control packets (which is highly energy inefficient due to the significant tail consumption after a cellular transmission~\cite{Balasubramanian:2009}).
\end{compactenum}

The above objectives involve some trade-offs, making it very challenging to satisfy all of them simultaneously. For instance, to maximize the traffic offload, we may consider a greedy approach, where the Content Server sends the content to users with the highest contact rates; however this would \begin{inparaenum}[(i)] \item deteriorate the fairness among users, and \item increase the signaling overhead to gather data on user mobility patterns\end{inparaenum}. Another approach may instead minimize the signaling overhead by injecting content as long as there is enough bandwidth available, avoiding thus any signaling; however, this will not maximize the traffic offload. In the following, we set the basic design guidelines of an approach that satisfies all these objectives.

\subsection{Basic design guidelines}

In order to satisfy the above objectives, a key decision of HYPE is how to deliver a certain piece of content (hereafter referred to as \emph{data chunk}) through the cellular network. In particular, this decision involves the selection of the nodes to which the data chunk is delivered via cellular, as well as the times when to perform these deliveries.

In HYPE, a data chunk is initially delivered to one or more users through the cellular network; additional copies may be injected later if needed. The decision of when to inject another copy of the chunk is driven by the number of users that have already received it. As long as the deadline has not expired, any user with a copy of the chunk will opportunistically transmit it to all the users it meets, that do not have the chunk. Finally, upon reaching the deadline of the content, the remaining users that have not yet received the chunk, download it from the cellular network;\footnote{An added advantage of this architecture is that the mobile nodes only need to keep the data chunks for forwarding until their deadline and no longer. The burden on the mobile nodes' buffers is thus kept very low.} this ensures that the \emph{delay guarantees} are met and thus we satisfy objective~(\ref{it:delay}) from Section~\ref{subsec:obj}.

In order to provide a good level of fairness among users, which is objective~(\ref{it:fair}), HYPE selects each of the seed nodes uniformly at random. Over the long term, this ensures that, on the one hand, all users have the same load in terms of cellular usage and, on the other hand, they also share fairly well the load incurred in opportunistic communications. This is confirmed by the simulation results presented in Section~\ref{sec-perfeval}, which show that HYPE provides a good level of fairness while paying a small price in terms of performance.\footnote{This is also supported by the results of \cite{Whitbeck2011}, which show that the difference in terms of performance between the random selection and other strategies is very small.}

\change{The approach sketched above meets objectives~(\ref{it:delay}) and~(\ref{it:fair}). In the following, we first present a model for the opportunistic dissemination of content injected by a cellular network. Based on this model, in Section~\ref{sec:opt_str} we derive the optimal strategy for the delivery of a single data chunk, that minimizes the load of the cellular network fulfilling objective~(\ref{it:offload}), and then we design an algorithm to implement this strategy, that incurs very low signaling overhead thus also satisfying objective~(\ref{it:signal}).}


\subsection{Model}\label{subsec:analysis}

\change{In order to derive the optimal strategy, with the above approach, for the delivery of data chunks through the cellular network, we need to determine:}
\begin{compactitem}
 \item The \emph{total number of copies} of the data chunk to be delivered by the cellular network. This is not trivial: for example, an overly conservative approach, that delivers too few copies before the deadline, may have the side-effect of overloading the cellular network with a large number of copies when the deadline expires.
 \item The \emph{optimal instants} for their delivery. The decision of when to deliver a copy of a data chunk through the cellular network is based on the current status of the network, which is given by the number of users that already have the chunk.
\end{compactitem}

\change{In the following, we model the opportunistic dissemination of content injected by a cellular network and analyze the load of the cellular network as a function of the strategy followed. Then, based on this analysis, in Section \ref{sec:opt_str} we obtain the optimal strategy, that minimizes the load of the cellular network for a given content deadline.}

Let $\mathcal{N}$ be a set of mobile nodes subscribed to the same content, with $N=|\mathcal{N}|$ the size of this set (total number of nodes). All nodes have access to the cellular network. Any two nodes also have the ability to setup pairwise bi-directional wireless links, when they are in each other's communication range (in \emph{contact}). Thus, opportunistic communication happens via the store-carry-forward method, through the sequences of intermittent contacts established by node mobility.

At time $0$, a data chunk is injected in the (opportunistic) network, i.e., copies of the chunk are pushed via the cellular interface to a small subset of $\mathcal{N}$, the seed nodes. Throughout the model description, we follow the epidemic dissemination of this chunk of content. We denote by $M(t)$ the number of mobile nodes holding the chunk at time $t$ (we refer to such nodes as ``infected''). The delivery deadline assigned to a data chunk is given by $T_c$ (its value depends on the mobile application's requirements).


\subsubsection{Opportunistic communication}
In the opportunistic phase of HYPE, data are exchanged only upon contacts in the network $\mathcal{N}$, therefore a mobility model based on contact patterns is sufficient for our analysis.


We assume every pair of nodes $(x,y)$ in the network $\mathcal{N}$ meets independently of other pairs, at exponentially distributed time intervals\footnote{Though all pairwise inter-contact rates may not always be exactly exponential (preliminary studies of traces~\cite{conan2007characterizing} suggested that this is true for subsets of node pairs only), the most in-depth and recent studies~\cite{Cai2009,Karagiannis2010} conclude that inter-contact time intervals do feature an exponential tail. This is supported by the recent results of Passarella et al.~\cite{Passarella2011}, which show that the non-exponential aggregated inter-contacts discovered in the preliminary trace studies~\cite{conan2007characterizing} can, in fact, be the result of exponentially distributed pairwise inter-contacts with different rates.} with rate $\beta_{xy} \geqslant 0$. Then, the opportunistic network $\mathcal{N}$ can be represented as a weighted contact graph using the $N\times N$ matrix $\mathbf{B} = \{\beta_{xy}\}$. We further assume that the inter-contact rates $\beta_{xy}$ are samples of a generic probability distribution $F(\beta):(0,\infty)\rightarrow[0,1]$ with known expectation $\mu_{\beta}$ (various distribution types for $F(\beta)$ and their effects on aggregated inter-contact times are investigated in~\cite{Passarella2011}). Additionally, we assume that the duration of a contact is negligible in comparison to the time between two consecutive contacts, and that the transmission of a single chunk is instantaneous in both the cellular and the opportunistic network.

The assumptions of exponential inter-contact and negligible contact duration are the norm in analytical work dealing with opportunistic networks~\cite{Groenevelt:2005,Neglia:2006,Picu2012a}. Studies based on looser assumptions (generic inter-contact models, non-zero contact duration) have, so far, only resulted in broad, qualitative conclusions (e.g., infinite vs. finite delay), while we aim at obtaining more concrete, quantitative results. In addition, all our simulations feature non-zero contact duration and some of them also have non-exponential inter-contact times, thus testing the applicability of our results outside the domain of these assumptions.

\begin{figure}[t]
\centering
\includegraphics[width=0.5\textwidth]{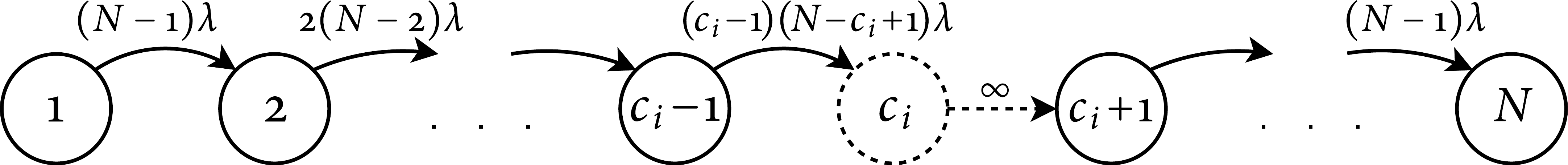}
\caption{\label{fig:mc_homo}Markov chain for HYPE communication, assuming \textbf{homogeneous node mobility}. Transitions can be caused either by (i) a contact between two nodes, or (ii) injection of the chunk to one node through the cellular network (instantaneous transition, represented with $\infty$ rate in the figure).}
\end{figure}

Epidemic dissemination in opportunistic networks is typically described with a pure-birth Markov chain, similar to the one in Fig.~\ref{fig:mc_homo} (slightly adapted from, e.g.,~\cite{Groenevelt:2005}). This type of chain only models the \emph{number of copies} of a chunk in the network $\mathcal{N}$ at any point in time, regardless of the specific nodes carrying those copies. This is only possible when considering node mobility to be entirely homogeneous (i.e., all node pairs meet at a unique rate: $\beta_{xy} = \lambda$ for all $x,y \in \mathcal{N}$), which allows all nodes to be treated as equivalent.

\begin{figure}[t]
\centering
\includegraphics[width=0.45\textwidth]{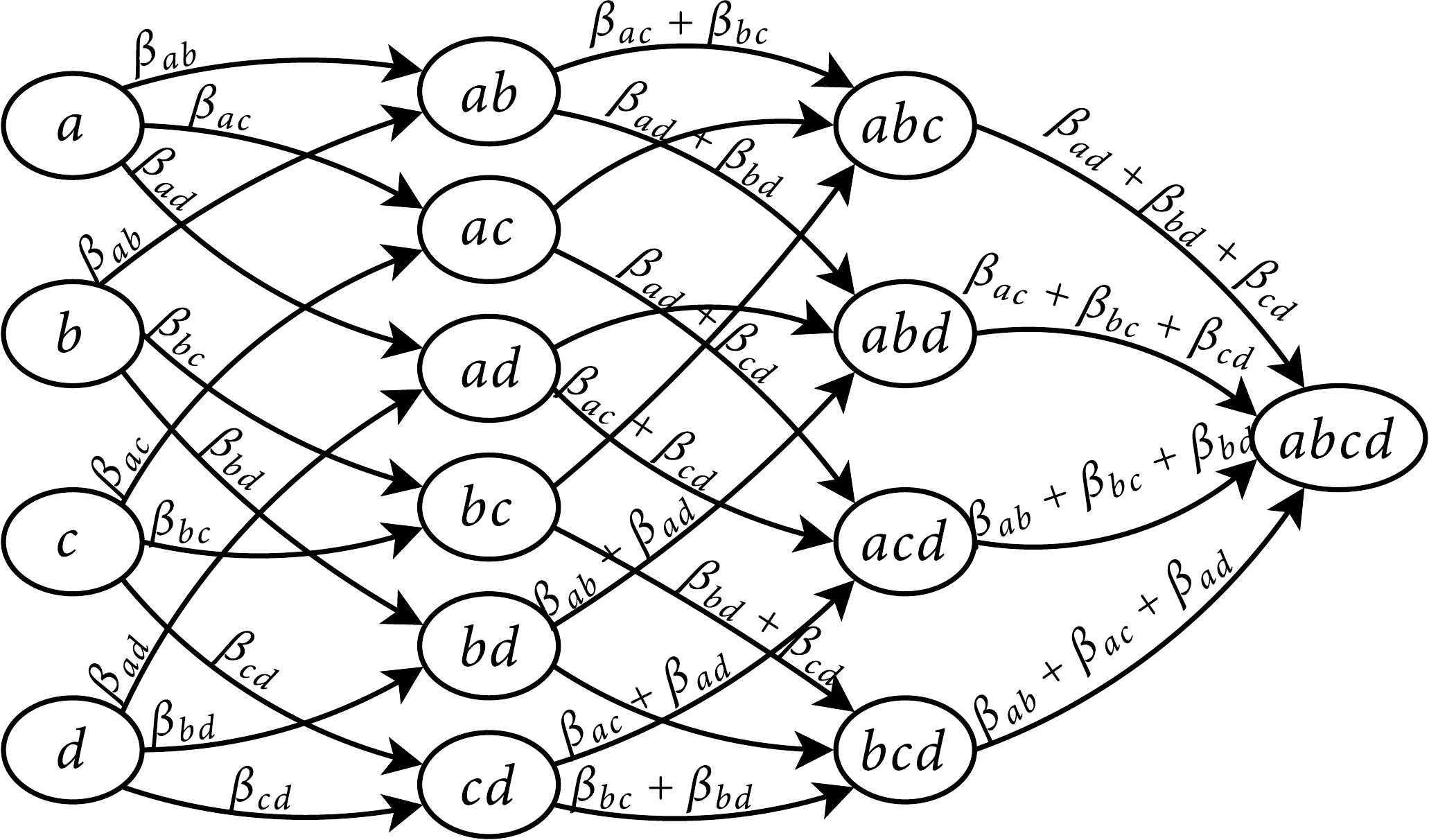}
\caption{\label{fig:mc_hetero}Markov chain for epidemic spreading, assuming \textbf{heterogeneous node mobility}. HYPE specific transitions (i.e., chunk injection by cellular) are left out for clarity. \change{This Markov chain is very complex and intractable for large scenarios; in Theorem~\ref{thm:mc_diff_eq} we can then reduce it to an equivalent Markov chain that is much simpler and for which we can derive a closed-form solution.}}
\end{figure}

However, as stated in the beginning of this subsection, we consider node mobility to be heterogeneous, with node pairs meeting at different rates $\beta_{xy}$ with $x,y \in \mathcal{N}$. In this case, not only the number of spread copies must be modeled, but also the \emph{specific nodes carrying those copies}. This results in more complex Markov chains, as illustrated in Fig.~\ref{fig:mc_hetero} for a $4$-node network $\mathcal{N} = \{a, b, c, d\}$.

Transition rates in Markov chains like the one shown in Fig.~\ref{fig:mc_hetero} depend on the nodes ``infected'' in each of the departure and the arriving states. For example, in Fig.~\ref{fig:mc_hetero}, the transition between state $a$ and state $ab$ can happen if node $a$ meets node $b$. Therefore, the transition time between these two states is exponential with rate given by the meeting rate of the $(a,b)$ node pair, $\beta_{ab}$. Similarly, the transition between state $ab$ and state $abc$ can happen if node $a$ meets node $c$, or if node $b$ meets node $c$ (whichever meeting happens first). Thus, the transition time for this transition is the minimum of two exponential variables with rates \change{$\beta_{ac}$} and $\beta_{bc}$. Since inter-contact times are exponential, this minimum is also exponential with rate \change{$\beta_{ac}+\beta_{bc}$}, as shown in Fig.~\ref{fig:mc_hetero}.

\subsubsection{Cellular communication}
The decision to deliver a copy of the chunk through the cellular network is based on the current dissemination level, i.e. the number of nodes that already have the chunk. We say that the HYPE process or its associated Markov chain (similar to Fig.~\ref{fig:mc_hetero}) is at \textbf{\emph{level}} $i$, when $i$ mobile nodes are infected, i.e., $M(t) = i$. Each level $i$ corresponds to a set of ${N\choose i}$ states $\{\mathbf{K}^{i}_{1},\mathbf{K}^{i}_{2},\ldots,\mathbf{K}^{i}_{{N\choose i}}\}$ in the Markov chain. For instance, in our $4$-node network from Fig.~\ref{fig:mc_hetero}, the HYPE process is at level $3$, when the chain is in any of the states $\mathbf{K}^{3}_{1} = abc$, $\mathbf{K}^{3}_{2} = abd$, $\mathbf{K}^{3}_{3} = acd$ or $\mathbf{K}^{3}_{4} = bcd$.

\add{The strategy to transmit copies of the chunk over the cellular network is given by the levels at which we inject a copy. We denote these levels by $C = \{c_1,c_2,\ldots,c_d\}$}: as soon as we reach one of these levels $c_i \in C$ before the deadline $T_c$, a copy of the chunk is sent to a randomly chosen node. With this, the transitions in the HYPE Markov chain can be caused either by: \begin{inparaenum}[(i)]\item a contact between two nodes (one infected, the other uninfected), which occurs at rates indicated in the previous subsection, or \item the injection of the chunk to one node through the cellular network\end{inparaenum}. The latter corresponds to an instantaneous transition (since the chain instantly ``jumps'' to a state of the next dissemination level), and is represented in Fig.~\ref{fig:mc_homo} with $\infty$ rate\footnote{Note that, for clarity, the Markov chain of Fig.~\ref{fig:mc_hetero} does not model transitions caused by chunk injection through the cellular network. This type of transition would be the same as in Fig.~\ref{fig:mc_homo} (i.e., $\infty$ rate).}. Finally, upon reaching the deadline $T_c$, the chunk is sent through the cellular network to those nodes that do not have the content by that time.


\subsection{Analysis}

Based on the above model, in the following, we analyze the load of the cellular network (which is the metric that we want to minimize) as a function of the strategy followed to inject content (which is given by $C  = \{c_1,c_2,\ldots,c_d\}$). The cellular network load corresponds to the number of copies delivered through the cellular network, which we denote by $D$. Let $p_i(t)=\mathbb{P}[M(t)=i]$ denote the probability of being at level $i$ at time $t$. Then, $D$ is given by:
\begin{equation}\label{eq-D}
D = \sum_{i = 1}^{N}{(d_i + {d}^*_i) p_i(T_c)}
\end{equation}
where $d_i$ is the number of deliveries through the cellular network that take place until level $i$ is reached ($d_i = \vert\{1,2,\ldots,i\} \cap C\vert$) and ${d}^*_i$ is the number of copies delivered upon reaching the deadline $T_c$, if it expires at level $i$ (${d}^*_i = N-i$).

In order to compute $p_i(T_c)$, we first analyze the case $C = \{c_1\}$\footnote{Note that $c_1$ must necessarily be equal to $0$.}, i.e., when we only inject one copy of the data chunk at the beginning and do not inject any other until we reach the deadline. Let $p_i^{c_1}(T_c)$ denote the probability that, in this case, the system is at level $i$ at time $T_c$. In order to compute $p_i^{c_i}(T_c)$, we model the transient solution of our Markov chain as shown in the following theorem. (The formal proofs of the theorems are provided in the Appendix.)

\begin{theorem}\label{thm:mc_diff_eq}
According to the HYPE Markov chain for heterogeneous mobility (similar to Fig.~\ref{fig:mc_hetero}), the process $\{M(t),t \ge 0\}$ is described by the following system of differential equations:
\begin{equation}
\begin{cases}
\frac{d}{d t} p_1^{c_1}(t) = - \lambda_1 p_1^{c_1}(t),& i= 1\\
\frac{d}{d t} p_i^{c_1}(t) = - \lambda_{i} p_{i}^{c_1}(t) + \lambda_{i-1} p_{i-1}^{c_1}(t), &  1 < i < N\\
\frac{d}{d t} p_N^{c_1}(t) = \lambda_{N-1} p_{N-1}^{c_1}(t),&  i= N
\end{cases}
\label{equ:birth_diff_equ}
\end{equation}
where $\lambda_{i} = i (N-i) \mu_{\beta}$. (Recall that $\mu_{\beta}$ is the known expectation of the generic probability distribution $F(\beta):(0,\infty)\rightarrow[0,1]$, from which the inter-contact rates describing our network are drawn: $\{\beta_{xy}\} = \mathbf{B}$.)
\end{theorem}

Theorem~\ref{thm:mc_diff_eq} has effectively reduced our complicated Markov chain for heterogeneous mobility back to a simpler Markov chain, like the one in Fig.~\ref{fig:mc_homo} (the $\lambda$ factor being replaced by $\mu_{\beta}$). In the simpler chain, each state represents a level of chunk dissemination (i.e., number of nodes holding a copy of the chunk). This is possible, as shown in the proof, thanks to the fact that our heterogeneous contact rates $\beta_{xy}$ are all drawn from the same distribution, $F(\beta):(0,\infty)\rightarrow[0,1]$, which means that all the states of a certain dissemination level $i$: $\{\mathbf{K}^{i}_{1},\mathbf{K}^{i}_{2},\ldots,\mathbf{K}^{i}_{{N\choose i}}\}$ are, in fact, statistically equivalent.

Applying the Laplace transform to the above differential equations, and taking into account that $p_i^{c_1}(0)=\delta_{i1}$, leads to
\begin{equation}
\begin{cases}
sP_{1}^{c_1}(s) =-\lambda_{1}P_1^{c_1}(s) + 1,& i= 1\\
sP_{i}^{c_1}(s) =-\lambda_i P_{i}^{c_1}(s) + \lambda_{i-1}P_{i-1}^{c_1}(s),&  1 < i < N\\
sP_{N}^{c_1}(s) = \lambda_{N-1}P_{N-1}^{c_1}(s),&  i= N
\end{cases}
\end{equation}
from which
\begin{equation}
\begin{cases}
P_{i}^{c_1}(s) = \displaystyle\frac{1}{s+\lambda_i} \prod_{j = 1}^{i-1}{\displaystyle\frac{\lambda_j}{s+\lambda_j}}, & i < N \\
P_{N}^{c_1}(s) = \displaystyle\frac{1}{s} \prod_{j = 1}^{N-1}{\displaystyle\frac{\lambda_j}{s+\lambda_j}},&  i= N
\end{cases}
\label{eq:laplace_probability}
\end{equation}

In case we deliver the data chunk through the cellular network at the levels $C = \{c_1, c_2, \ldots, c_d\}$, then the transitions corresponding to those levels are instantaneous, and the Laplace transforms of the probabilities
$P_i(s)$ are computed as:
\begin{equation}\label{eq-laplace}
P_{i}(s)=
\begin{cases}
\displaystyle\frac{1}{s+\lambda_i} \prod_{j \in S_{i-1}}{\displaystyle\frac{\lambda_j}{s+\lambda_j}}, & i < N,\;i \notin C\\
0, & i < N,\;i \in C\\
\displaystyle\frac{1}{s} \prod_{j \in S_{N-1}}{\displaystyle\frac{\lambda_j}{s+\lambda_j}},&  i= N
\end{cases}
\end{equation}
where $S_{i-1}$ is the set of levels up to level $i-1$, without including those that belong to set $C$, i.e., $S_{i-1} = \{1,2,\ldots,i-1\} \setminus (\{1,2,\ldots,i-1\} \cap C)$. For the levels $i \in C$, we simply have $P^C_{i}(s) = 0$, since we will never be at these levels.

From Eq.~(\ref{eq-laplace}), we can obtain a closed-form expression for the probabilities $p_i(t)$ as follows. The polynomial $P_{i}(s)$ is characterized by first and second order poles which have all negative real values. Let $\{s=-\lambda_n\}$ be the poles of $P_{i}(s)$. Then, $p_i(t)$ for $i < N, i \notin C$ is computed as:
\begin{equation}
p_i(t) = \left(\prod_{j \in S_{i-1}} \lambda_j\right)
\sum_{\{s=-\lambda_n\}}Res\left(\frac{e^{st}}{\prod_{j \in S_{i}}(\lambda_j+s)}\right)
\label{equ:birth_pk_res}
\end{equation}
where $Res$ indicates the residue, which is given by:
\begin{eqnarray}
\hspace{-1cm}&\underset{s=-\lambda_n}{Res}\left(\displaystyle\frac{e^{st}}{\prod\limits_{j \in S_{i}}(\lambda_j+s)}\right)= \nonumber\\\nonumber
\hspace{-0.4cm}&\begin{cases}
  \displaystyle\frac{e^{-\lambda_nt}}{\prod\limits_{\substack{j \in S_{i}\\j\neq n}}(\lambda_j-\lambda_n)}, & -\lambda_n \mbox{ is a $1^\text{st}$ order pole}\\
  \displaystyle\frac{e^{-\lambda_nt}\left[t-\sum\limits_{\substack{j \in S_{i}\\\lambda_r\neq\lambda_n}}\displaystyle\frac{1}{(\lambda_r-\lambda_n)}\right]}{\prod\limits_{\substack{j \in S_{i}\\\lambda_j\neq\lambda_n}}{(\lambda_j-\lambda_n)}}, & -\lambda_n \mbox{ is a $2^\text{nd}$ order pole}
 \end{cases}
\end{eqnarray}

Additionally, for $i < N, i \in C$ we have $p_i(t)=0$, and for $i= N$, $p_{N}(t)=1-\sum_{i=1}^{N-1} p_k(t)$. 

By evaluating $p_{i}(t)$ at time $t = T_c$ and applying Eq.~(\ref{eq-D}), we can compute the average number of deliveries over the cellular network, $D$. 

\section{\change{Optimal Strategy and Adaptive Algorithm}}\label{sec:opt_str}

\change{In this section, we first leverage on the above model to determine the optimal strategy for the delivery of data chunk, and then we design an adaptive algorithm to implement this strategy.}

\subsection{\change{Optimal strategy analysis}}

Our goal is to find the best strategy $C  = \{c_1,c_2,\ldots,c_d\}$ for injecting chunk copies over the cellular network, that minimizes the total load $D$ of the cellular network while meeting the content's deadline $T_c$. To solve this optimization problem, we proceed along the following two steps:
\begin{compactenum}
\item We show that the optimal strategy is to deliver the content through the cellular network only at the beginning and at the end of the data chunk's \emph{period}, and never in-between. The data chunk's \emph{period} is defined as the interval between $t = 0$ (when we first start distributing the content) and $t = T_c$ (when the content's deadline expires).
\item We obtain the optimal number of copies of the chunk to be delivered at the beginning of the period such that the average load of the cellular network, $D$, is minimized.
\end{compactenum}

The following theorem addresses the first step.
\begin{theorem}\label{thm:opt_when}
In the optimal strategy, the data chunk is delivered through the cellular network to $d$ seed nodes at time $t = 0$, and to the nodes that do not have the content by the deadline at time $t = T_c$.
\end{theorem}

According to Theorem~\ref{thm:opt_when}, the optimal strategy is to: \begin{inparaenum}[(i)] \item deliver a number of copies through the cellular network at the beginning of the period, \item wait until the deadline \emph{without} delivering any additional copy, and \item deliver a copy of the chunk to the mobile nodes missing the content at the end of the period\end{inparaenum}.

The intuition behind this result is as follows. When few users have the content, information spreads slowly, since it is unlikely that a meeting between two nodes involves one of the few that have already the content. Similarly, information spreads slowly when many users have the content, as a meeting involving a node that does not yet have the content is improbable.

The strategy given by Theorem~\ref{thm:opt_when} avoids the above situations by delivering a number of chunk copies through cellular communication at the beginning (when few users have the content) and at the end (where few users miss the content). As a result, the strategy lets the content disseminate through opportunistic communication when the expected speed of dissemination is higher, which allows to minimize the average load of the cellular network.

\change{The second challenge in deriving the optimal strategy is to compute the optimal number of copies of the chunk to be delivered at the beginning of the period, which we denote by $d$. To that end, the following proposition defines the notion of gain and computes it:}

\begin{proposition}\label{prop:opt_number} Let us define $G_d$ as the gain resulting from sending the $(d+1)^{th}$ chunk of chunk copy at the beginning of the period (i.e., $G_d = D_{d} - D_{d+1}$, where $D_{d+1}$ and $D_{d}$ are the values of $D$ when we deliver $d+1$ and $d$ copies at the beginning, respectively). Then, $G_d$ can be computed from the following equation:
\begin{equation}\label{eq-th3}
G_d = \sum_{j = d}^{N-1}{\frac{\lambda_j}{\lambda_{d}}p_{j}^d(T_c)} - 1,
\end{equation}
\end{proposition}

\change{Building on the above notion of $G_d$, the following theorem provides the optimal point of operation:}
\begin{theorem}\label{thm:opt_number}
The optimal value of $d$ is the one that satisfies $G_d  = 0$.
\end{theorem}

The rationale behind the above theorem is as follows. When $G_d > 0$, by sending one additional copy at the beginning, we save more than one copy at the end of the period and hence obtain a gain. Conversely, when $G_d < 0$, we do not benefit from increasing $d$. The proof shows that $G_d$ is a strictly decreasing function of $d$, which implies that, to find the optimal point of operation, we need to increase $d$ as long as $G_d > 0$ and stop when we reach $G_d = 0$ (after this point, $G_d < 0$ and further increasing $d$ yields a loss).



\subsection{Adaptive algorithm for optimal delivery}\label{subsec:adapt_algo}

\change{While the previous section addressed the delivery of a single data chunk, in this section we focus on the delivery of the entire content, e.g., a flow of road traffic updates, news feeds or a streaming sequence. We consider that the distribution of content in mobile applications is typically performed by independently delivering different pieces of content in a sequence of data chunks. For instance, a streaming content of $800$\,MB may be divided into a sequence of chunks of $1$\,MB.} When delivering chunks in sequence, we need to adapt to the system dynamics. For instance, inter-contact time statistics may vary depending on the time of the day~\cite{chaintreau}, which means that the optimal $d$ value obtained by Theorem~\ref{thm:opt_number} needs to be adapted accordingly. Similarly, the number of mobile nodes $N$ subscribed to the content may change with time, e.g., based on the content popularity.

To address the above issues, we design an adaptive algorithm based on control theory, that adjusts the number of chunk copies $d$ delivered at the beginning of each period to the behavior observed in previous \emph{rounds} (hereafter we refer to the sequence of periods as rounds). \change{For instance, in the example above we would have a total of $800$ rounds.} In the following, we first present the basic design guidelines of our adaptive algorithm. Building on these guidelines, we then design our system based on control theory. Finally, we conduct an analysis of the system to guarantee its stability and ensure good response times.

\subsection{Adaptive algorithm basics}\label{subsec:adapt_algo_basics}

In order to devise an adaptive algorithm that drives the system to optimality, we first need to identify \emph{which variable} we should monitor and \emph{what value} this variable should take in optimal operation. To do this, we build on the results of the previous section to design an algorithm that: \begin{inparaenum}[(i)]\item monitors \emph{how many additional infected nodes} we would have at the end of a round, if we injected one extra copy at the beginning of that round; and \item drives the system to optimality by increasing or decreasing $d$ depending on whether this number is above or below its optimal value. \end{inparaenum}

To efficiently monitor the number of additional infected nodes, we apply the following reasoning. According to Theorem~\ref{thm:opt_number}, in optimal operation, one extra delivery at the beginning of a round leads to one additional infected node at the end of that round. If we focus on a single copy of the chunk delivered over the cellular network and consider it as the extra delivery, the nodes that would receive the content due to this one extra delivery are those that \emph{\add{received this specific copy and could not have received the chunk from any other source}}. Since this holds for each of the $d$ copies delivered over the cellular network, in optimal operation there are on average a total of $d$ nodes at the end of the round, which received \add{the chunk from one source and could not have received it from any other source}. Our algorithm focuses on this aggregate behavior of the $d$ deliveries rather than on a single copy, as this provides more accurate information about the epidemic dissemination of the data chunk.

\begin{figure}[t]
\centering
 \includegraphics[width=0.49\textwidth]{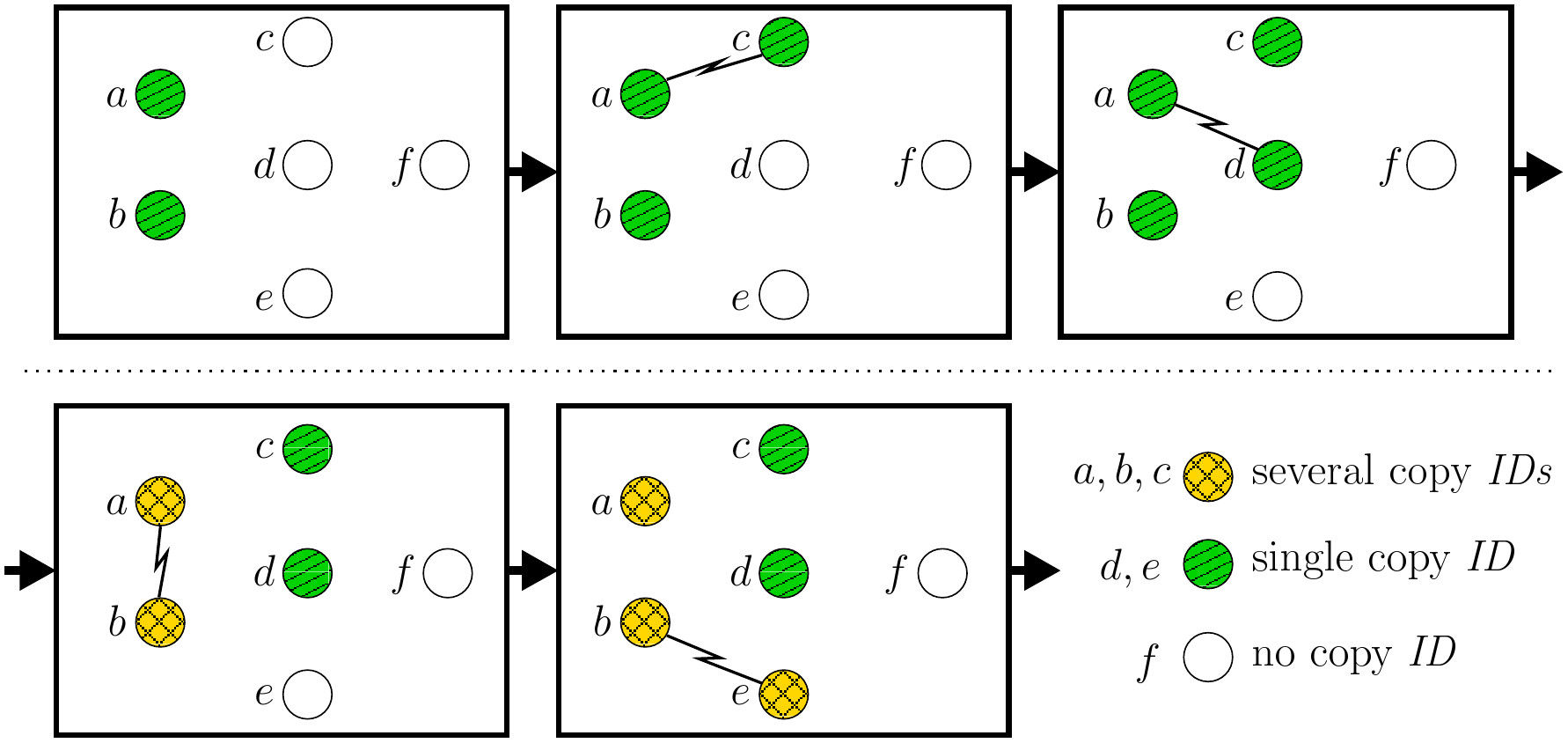}
 \caption{\change{Example of chunk dissemination in optimal operation}. Node $a$ and $b$ receive a copy of the chunk from the Content Server ($d=2$). At the end of the round, there are two nodes with a single copy ID, that is, $s=2$.}
  \label{fig:dissemination}
\end{figure}

Based on this, each round of the adaptive algorithm proceeds as follows (see Fig.~\ref{fig:dissemination} for an example):
\begin{compactenum}
\item Initially, copies of the data chunk are transmitted to a random set of $d$ seed nodes over the cellular network. Each of the copies is marked with a different \emph{ID} that uniquely identifies the source of the copy. 
\item When a node that does not have the chunk receives it from another node opportunistically, it records the \emph{ID} of the copy received.
\item If two nodes that have copies with different \emph{ID}s meet, they mark this event, to record that they could have potentially received the chunk from different sources.\footnote{Note that, for a node with ``several copy \emph{ID}s'', we only mark the event and do not keep the \emph{ID}s of the copies, since (i) we are only interested in signaling the number of nodes with a single copy \emph{ID}, and (ii) this leads to more efficient operation, requiring fewer communications and less protocol overhead.} We say that such nodes have ``several copy \emph{ID}s'', while those that keep only one \emph{ID} have a ``single copy \emph{ID}''.
\item If a node who does not have any copy or has a single copy \emph{ID} meets with another node who recorded the ``several copy \emph{ID}s'' event, the first node also marks its copy with the ``several copy \emph{ID}s'' mark.
\item At the end of the round, the nodes whose chunk comes from a single source (i.e., no ``several copy \emph{ID}s'' mark) send a signal to the Content Server.
\end{compactenum}

By running the above algorithm, we count the number of nodes whose copy of the chunk comes from a single source (i.e., who have a single copy \emph{ID} at the end of the round), which we denote by $s$. 
\add{As argued at the beginning of this section, in optimal operation this number is equal to the number of seed nodes. This implies that, at this operating point, \emph{the number of data chunks injected through the cellular network at the beginning of the round} is equal\change{, in expectation,} to \emph{the number of signals received at the end of the round}, i.e., $s = d$}.

A key feature of the above algorithm design is that it does not require gathering any complex statistics on the network, such as the behavior of the mobile nodes, their mobility or social patterns, or their contact rates. Instead, we just need to keep track of the number of chunks injected at the beginning of each round and the signals received at the end, and this is sufficient to drive the system to optimal operation. As a result, the proposed algorithm involves very reduced signaling overhead, which fulfills one of the objectives that we had identified in Section~\ref{subsec:obj}, namely objective~(\ref{it:signal}).

\subsection{System design}

Based on the above design guidelines, our adaptive algorithm should \begin{inparaenum}[(i)] \item monitor the number of signals received at the end of each round, and \item drive the system to the point of operation where this value is equal to the number of copies injected at the beginning of the round\end{inparaenum}. To do this, in this paper we rely on control theory, which provides the theoretical basis for monitoring a given variable (the \emph{output signal} in control theory terminology) and driving it to some desired value (the \emph{reference signal}).

\begin{figure}[t]
\centering
 \includegraphics[width=0.45\textwidth]{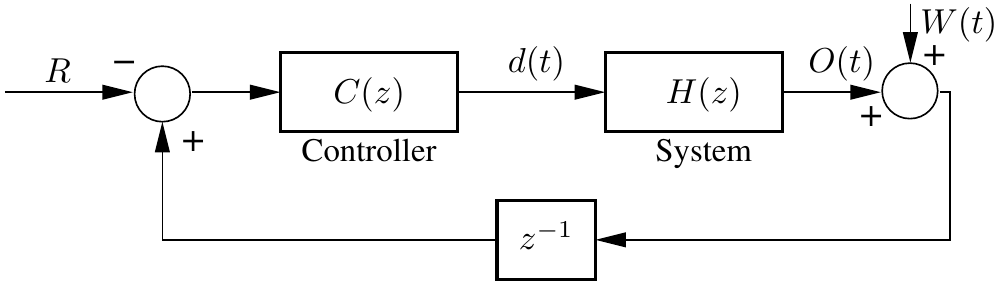}
 \caption{Our system is composed by two modules: the controlled system $H(z)$, that models the behavior of HYPE, and the PI controller $C(z)$, that drives
the controlled system to the optimal point of operation.}
  \label{fig:controller}
\end{figure}

Following a control theoretic design, we propose the system depicted in Fig.~\ref{fig:controller}. This system is composed from a controller $C(z)$, which is the adaptive algorithm that controls the chunk delivery, and the controlled system $H(z)$, which represents the HYPE network. Furthermore, the component $z^{-1}$ provides the delay in the feedback-loop (to account for the fact that the $d$ value used in the current round is computed from the behavior observed in the previous round). For the controller, we have decided to use a Proportional-Integral (PI), because of its simplicity and the fact that it guarantees zero error in the steady-state. The $z$ transform of the PI controller is given by:
\begin{equation}
C(z) = K_p + \frac{K_i}{z-1}
\end{equation}
where $K_p$ and $K_i$ are the parameters of the controller.

Here, the variable that we want to optimally adjust is the number of deliveries at the beginning of the round (i.e., $d$). Following classical control theory~\cite{astrom}, this variable is the \emph{control signal} provided by the controller. In each round, the controller monitors the system behavior (and in particular the output signal, which we will define later), given the value $d$ that is currently used. Based on this behavior, it decides whether to increase or decrease $d$ in the next round, in order to drive the output signal to the reference signal.

A key aspect of the system design is the definition of the output and reference signals. On the one hand, we need to enforce that by driving the output signal to the reference signal, we bring the system to the optimal point of operation. On the other hand, we also need to ensure that the reference signal is a constant value that does not depend on variable parameters, such as the number of nodes or the contact rates.

Following the arguments exposed in Section~\ref{subsec:adapt_algo_basics}, we design the output signal $O(t)$ and the reference signal $R$ of our controller as follows:
\begin{equation}
\begin{cases}
O(t) =s(t) - d(t)\\
R = 0
\end{cases}
\end{equation}
where $d(t)$ is the number of deliveries at the beginning of a given round $t$, and $s(t)$ is the number of signals received at the end of this round. Note that, with the above output and reference signals, by driving $O(t)$ to $R$ we bring the system to the point of operation given by $s = d$, which, as discussed previously, corresponds to the optimal point of operation. \change{Following classical control theory, we represent the randomness of the system by adding some noise $W(t)$ to the output signal, as shown by Fig.~\ref{fig:controller}.}

\subsection{Control theoretic analysis}

The behavior of the proposed system (in terms of stability and response time) depends on the parameters of the controller $C(z)$, namely $K_p$ and $K_i$. In the following, we conduct a control theoretic analysis of the system and, based on this analysis, calculate the setting of these parameters.
\change{Note that this analysis guarantees that the algorithm quickly converges to the desired point of operation and remains stable at that point.}

In order to analyze our system from a control theoretic standpoint, we need to characterize the HYPE network with a transfer function $H(z)$ that takes $d$ as input and provides $s-d$ as output. \change{In order to derive $H(z)$, we proceed as follows. According to the definition given in Proposition~\ref{prop:opt_number}, $G_d$ is the gain resulting from sending an extra copy of the chunk. In one round, by sending one extra copy of the chunk at the beginning, there are on average $s/d$ additional nodes that have the chunk at the end. Indeed, $s$ is the total number of nodes that receive the chunk from only one of the $d$ initial seed nodes, which means that on average each seed node contributes with $s/d$ to this number.} This yields to:
\begin{equation}
G_d = s/d - 1,
\end{equation}
from which:
\begin{equation}\label{eq-d*d}
s-d = G_d d.
\end{equation}
The above provides a nonlinear relationship between $d$ and $s-d$, since $G_d$ (given by Eq.~(\ref{eq-th3})) is a non-linear function of $d$. To express this relationship as a transfer function $H(z)$, we linearize it at the optimal point of operation.\footnote{This linearization provides a good approximation of the behavior of the system when it suffers small perturbations around the stable point of operation~\cite{kalman}. \change{Note that the approximation only affects the transient analysis and not the analysis of the stable point of operation at which the system is brought by the algorithm.}} Then, we study the linearized model and ensure its stability through appropriate choice of parameters. Note that the stability of the linearized model guarantees that our system is locally stable.\footnote{A similar approach was used in~\cite{misra:infocom2001} to analyze the Random Early Detection (RED) scheme from a control theoretic standpoint.}

To obtain the linearized model, we approximate the perturbations suffered by $s-d$ at the optimal point of operation, $\Delta(s-d)$, as a linear function of the perturbations suffered by $d$, $\Delta d$,
\begin{equation}
\Delta(s-d) \approx \frac{\partial (s-d)}{\partial d} \Delta d,
\end{equation}
which gives the following transfer function for the linearized system:
\begin{equation}
H(z) = \frac{\partial (s-d)}{\partial d}.
\end{equation}

Combining the above with Eq.~(\ref{eq-d*d}), we obtain the following expression for $H(z)$:
\begin{equation}
H(z) = \frac{\partial (s-d)}{\partial d} = G_d + d \frac{\partial G_d}{\partial d}.
\end{equation}

Evaluating $H(z)$ at the optimal point of operation ($G_d = 0$) yields:
\begin{equation}
H(z) = d \frac{\partial G_d}{\partial d}.
\end{equation}

To calculate the above derivative, we approximate $\lambda_i$ (given by $\lambda_i = i (N-i)\mu_{\beta}$) by its first order Taylor polynomial evaluated at level $i = \hat{d}$, where $\hat{d}$ is the average value of $i$ at time $T_c$ (i.e., the average number of nodes that have the chunk at the deadline). Since the Taylor polynomial provides an accurate approximation for small perturbations around $\hat{d}$, and the number of nodes that have the chunk at time $T_c$ is distributed around this value, we argue that this approximation leads to accurate results. The first order Taylor polynomial for $\lambda_i$ at $i = \hat{d}$ is:
\begin{equation}
\lambda_i \approx \lambda_{\hat{d}} - (i - \hat{d})(2 \hat{d} - N)\mu_{\beta}.
\end{equation}

Substituting this into Eq. (\ref{eq-th3}) yields
\begin{align}
G_d & = \frac{1}{\lambda_d}\sum_{i=1}^N{p_i^d(T_c)\left(\lambda_{\hat{d}} - (i - \hat{d})(2\hat{d}-N)\lambda \right)} - 1 \nonumber\\
& = \frac{\lambda_{\hat{d}}}{\lambda_d} - 1 = \frac{\hat{d} (N-\hat{d})\mu_{\beta}}{d (N-d)\mu_{\beta}} - 1
\end{align}

Since at the optimal point of operation we have $G_d = 0$, this implies that (at this operating point) $d = \hat{d}$. Moreover, from Theorem~\ref{thm:opt_number} we have that, when operating at the optimal point, if we deliver one additional copy at the beginning (i.e., increase $d$ by one unit), this leads to one additional node with the chunk at the end (i.e., $\hat{d}$ also increases by one unit). Therefore, at the optimal operating point we also have $\partial \hat{d}/\partial d = 1$. Accounting for all of this when performing the partial derivative of $G_d$ yields:
\begin{equation}
\frac{\partial G_d}{\partial d} = \frac{2(2d-N)}{d(N-d)},
\end{equation}
from which:
\begin{equation}
H(z) = d\frac{\partial G_d}{\partial d} = - \frac{2(N-2d)}{N-d}.
\end{equation}

Having obtained the transfer function of our HYPE network, we finally address the configuration of the controller parameters $K_p$ and $K_i$, that will ensure a good trade-off between our system's stability and response time. To this end, we apply the Ziegler-Nichols rules~\cite{franklin}, which have been designed for this purpose. According to these rules, we first obtain the $K_p$ value that leads to instability when $K_i = 0$; this value is denoted by $K_u$. We also calculate the oscillation time $T_i$ under these conditions. Once the $K_u$ and $T_i$ values have been derived, $K_p$ and $K_i$ are configured as follows:
\begin{equation}\label{eq-kukp}
K_p = 0.4 K_{u}, \ \ K_i = \frac{K_{p}}{0.85 T_i}.
\end{equation}

Let us start by computing $K_u$, i.e., the $K_p$ value that ensures stability when $K_i = 0$. From control theory~\cite{astrom}, we have that the system is stable as long as the absolute value of the closed-loop gain is smaller than $1$. The closed-loop transfer function $T(z)$ of the system depicted in Fig.~\ref{fig:controller} is given by:
\begin{equation}
T(z) = \frac{- H(z) C (z)}{1-z^{-1}H(z) C (z)}.
\end{equation}

To ensure that the closed-loop gain of the above transfer function is smaller than $1$, we need to impose $\vert H(z) C (z) \vert < 1$. Doing this for $K_i = 0$ yields:
\begin{equation}
\vert H(z) C (z) \vert = \left| -\frac{2(N-2d)}{N-d} K_p \right|  < 1.
\end{equation}

The above inequality gives the following upper bound for $K_p$, at which the system turns unstable: 
\begin{equation}
K_p < \frac{N-d}{2(N-2d)}.
\label{eq:inequality}
\end{equation}

We want to ensure that the system is stable independently of $N$ and $d$, that is, the above inequality holds for any $N$ and $d$ values. Since the smallest possible value that the right-hand side of Eq.~(\ref{eq:inequality}) can take is $1/2$ (when $d \rightarrow 0$), the system is guaranteed to be stable as long as $K_p < 1/2$, and may turn unstable when $K_p$ exceeds this value. Accordingly, we set $K_u = 1/2$. Furthermore, when the system becomes unstable, the control signal $d$ may change its sign up to every round, yielding an oscillation period of two rounds, which gives $T_i = 2$. With these $K_u$ and $T_i$ values, we set $K_p$ and $K_i$ following Eq.~(\ref{eq-kukp}),
\begin{equation}\label{eq-kp}
K_p = \frac{0.4}{2}, \ \ K_i = \frac{0.4}{2 \cdot 2 \cdot 0.85},
\end{equation}
which terminates the configuration of the PI controller.

While the Ziegler-Nichols rules aim at providing a good trade-off between stability and response time, they are heuristic in nature and thus do not guarantee the stability of the system. The following theorem proves that the system is stable with the proposed configuration.

\begin{theorem} The HYPE control system is stable for $K_p = 0.2$ and $K_i = 0.4/3.4$.\newline
\end{theorem}



\section{Performance Evaluation}\label{sec-perfeval}

In this section, we evaluate HYPE for a wide range of scenarios, including several instances of a heterogeneous mobility model, as well as real-world mobility traces. We show that:
\begin{compactitem}
 \item The analytical model provides very accurate results.
 \item The optimal strategy for data chunk delivery effectively minimizes the load incurred in the cellular network.
 \item The proposed adaptive algorithm is stable and quickly converges to optimal operation.
 \item HYPE outperforms previously proposed heuristics in terms of the cellular load, signaling load and fairness among users.
\end{compactitem}

\change{From the four design objectives introduced in Section~\ref{subsec:obj}, our evaluation focuses on the traffic offload, fairness and signaling overhead. Note that, since the delay guarantees are satisfied by design, we meet the objective on the delay.}

\paragraph{Simulation setting}

To evaluate the performance of HYPE, we use both real mobility traces and a heterogeneous mobility model. For the evaluation with real mobility traces, we select the contact traces collected in the Haggle project for $4$ days during Infocom 2006~\cite{chaintreau}, and the GPS location traces of San Francisco taxicabs,\footnote{We assume two taxicabs are in contact when they are within $100$ meters of each other.} collected through the Cabspotting project~\cite{epfl-mobility-2009-02-24}. The number of users for the Infocom 2006 and San Francisco traces are $78$ and $536$, respectively.

As for the heterogeneous mobility model, we generate contacts as follows. For any given node pair $(x,y)$, the pairwise inter-contact times are exponentially distributed with rate $\beta_{xy}$. The pairwise contact rates, $\beta_{xy}$, are drawn from a Pareto distribution\footnote{Under these conditions, the tail of the aggregate inter-contact times decays as a power law with exponential cut-off~\cite{Passarella2011}, as observed in traces, in~\cite{Karagiannis2010}.} with mean $\mu_{\beta}$ (which determines the average frequency of the contacts) and standard deviation $\sigma$ (which determines the level of heterogeneity). 
To account for sparser scenarios, we also run some experiments where a node pair has a probability $p > 0$ of never meeting, i.e., $\beta_{xy} = 0$ (otherwise the inter-contact rate for the pair $\beta_{xy}$ is drawn as above).
%
In addition, we generate contact durations $\delta$ from a Pareto distribution with parameter $\alpha = 2$, as observed in~\cite{chaintreau2005pocket}. Following the findings in~\cite{conan2007characterizing}, we choose the average contact rate $\mu_\beta$ and the average contact duration $\mathbb{E}[\delta]$ values such that $1/(\mu_\beta \cdot \mathbb{E}[\delta])$ is between $100$ and $1000$.


In all the simulations, we set the throughput of the cellular communication to one mobile node equal to $600$\,kb/s~\cite{Balasubramanian} and the bandwidth of opportunistic communication to $20$\,Mb/s. All the results given in this section are provided with $95$\% confidence intervals below $0.1$\%. 

\paragraph{Baseline scenarios}
For the heterogeneous mobility model, we use the following four baseline scenarios:  
\begin{compactitem}
\item \textbf{streaming}: $N = 100$, mean contact rate $\mu_\beta$ = $13$ contacts/pair/day~\cite{chaintreau} and $\sigma = 0.58 \cdot \mu_\beta$, Pareto-distributed contact duration $\mathbb{E}[\delta]$=$66.46$\,s, $T_c$ = $120$\,s~\cite{Sentinelli} and chunk size $L = 1$\,MB,
\item \textbf{road traffic update}: $N = 1000$, mean contact rate $\mu_\beta$  = $1.2$ contacts/pair/day and $\sigma = 1.5 \cdot \mu_\beta$, Pareto-distributed contact duration $\mathbb{E}[\delta]$ = $72$\,s, $T_c$ = $600$\,s, $L =1$\,MB~\cite{Whitbeck2011},
\item \textbf{news feed}: $N = 100$, mean contact rate $\mu_\beta$  = $0.69$ contacts/pair/day~\cite{chaintreau} and $\sigma = 2 \cdot \mu_\beta$, Pareto-distributed contact duration $\mathbb{E}[\delta]$= $125$\,s, $T_c$ = $3600$\,s~\cite{Sentinelli}, $L = 0.5$\,MB,
\item \textbf{social data}: $N = 50$, mean contact rate $\mu_\beta$  = $3.5$ contacts/pair/day~\cite{chaintreau} and $\sigma = \mu_\beta$, Pareto-distributed contact duration $\mathbb{E}[\delta]$= $164$\,s, $T_c$ = $900$\,s, $L =4$\,KB.
\end{compactitem}

\subsection{Validation of the model}

\begin{figure}[t]
\centering
 \includegraphics[width=0.49\textwidth]{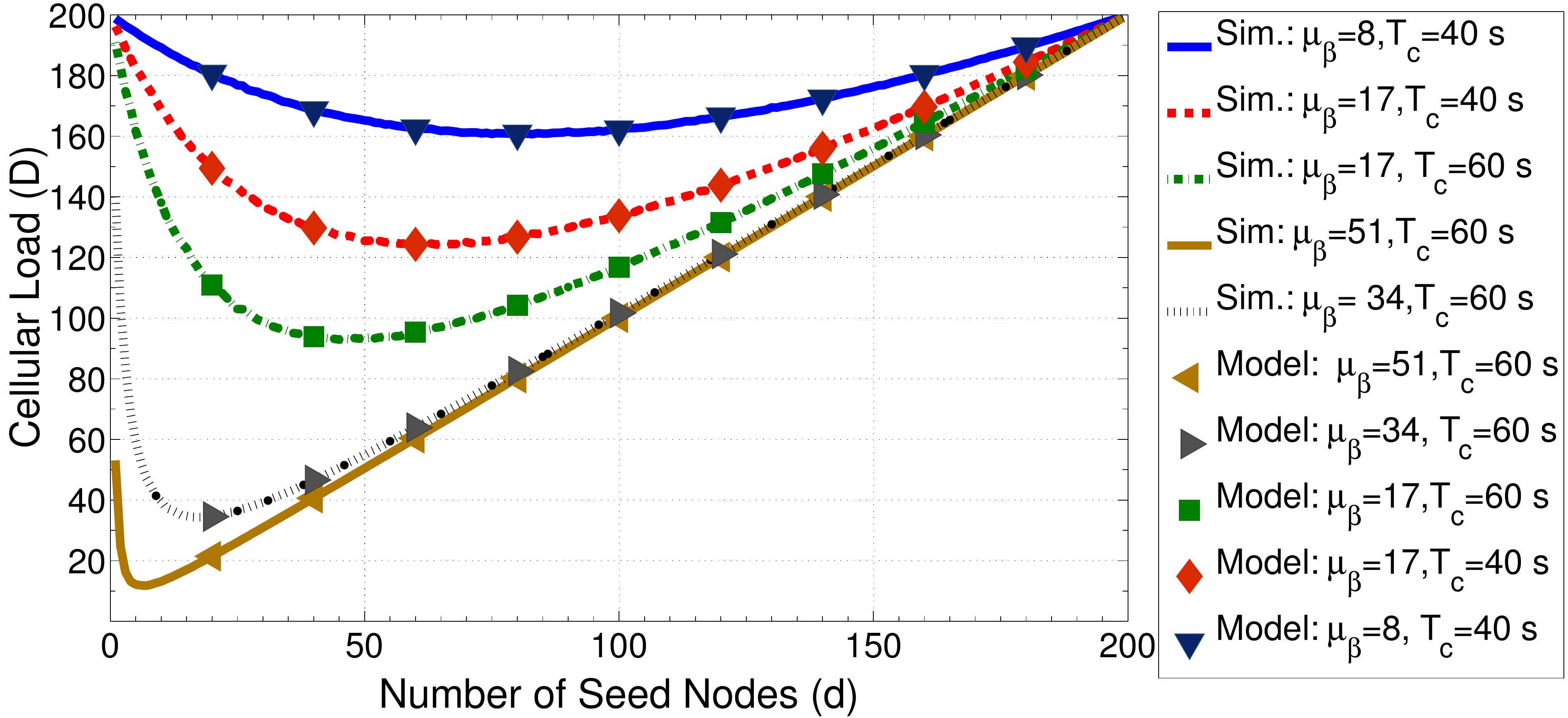}
 \caption{The analytical model provides very accurate results for different settings ($\mu_\beta$ is given in contacts/pair/day).}
  \label{fig:validation}
\end{figure}

In order to validate the analysis conducted in Section~\ref{sec:arch}, we evaluate the total load incurred in the cellular network ($D$) as a function of the strategy followed (which is given by the number of copies of the data chunk delivered at the beginning of a round, $d$). The results obtained are depicted in Fig.~\ref{fig:validation} for a scenario with $N = 200$, $\sigma=0.04$ contacts/pair/day, and different values of $T_c$ (in seconds) and $\mu_\beta$ (in contacts/pair/day). We observe that the analytical results follow very closely those resulting from simulations, which validates the accuracy of our analysis. \change{We further observe that, as pointed out in Section~\ref{sec:opt_str}, performance degrades for smaller and larger values of $d$, since when either too few or too many nodes have the content, information spreads slowly. The figure finally shows that  -- given $\mu_b=17$ --  a smaller $T_c$ (of $40$\,s) causes a higher load of cellular network than a larger one (of $60$\,s).}


\begin{figure}[t]
\centering
 \includegraphics[width=0.49\textwidth]{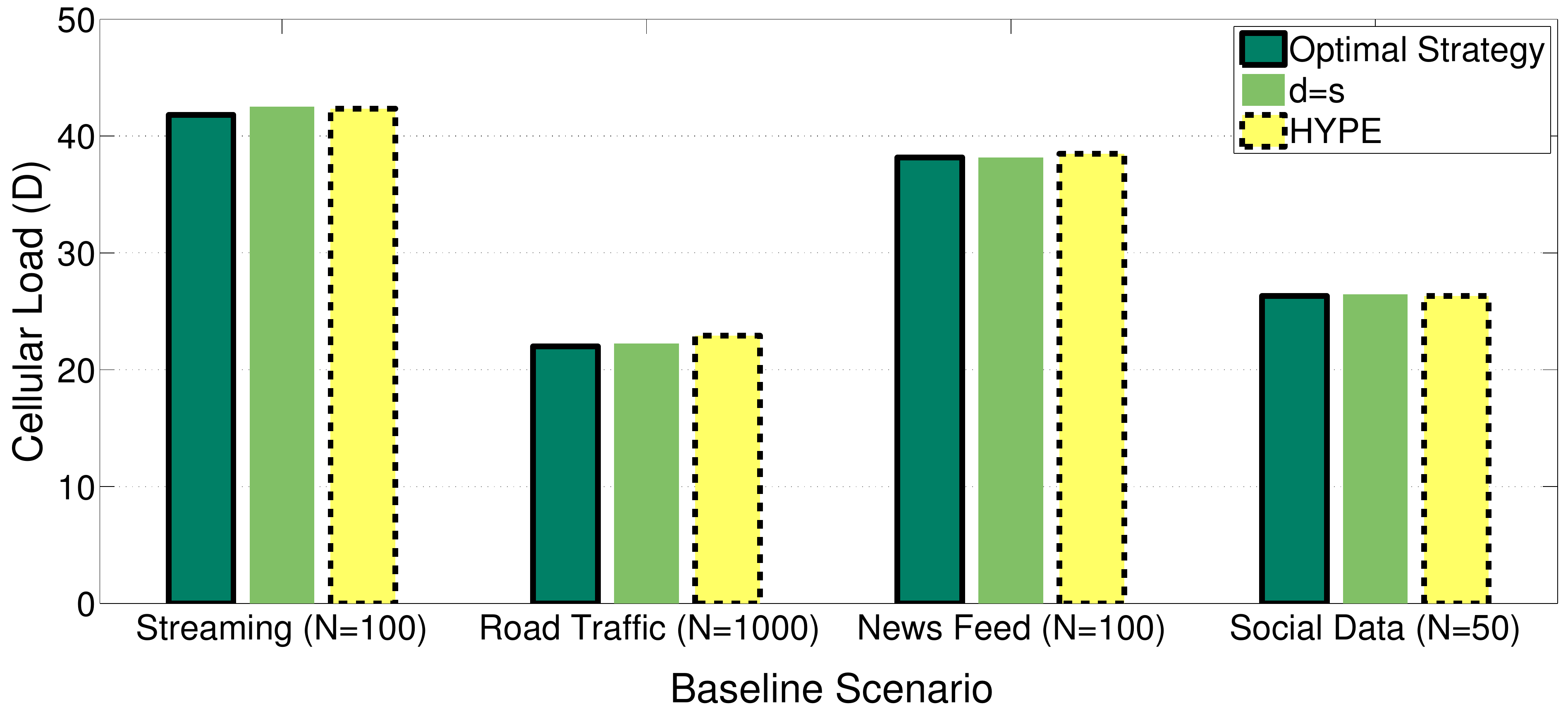}
 \caption{Validation of the optimal strategy for the four baseline scenarios.}
  \label{fig:bar_gain}
\end{figure}

\subsection{\change{Performance gain and validation of the optimal strategy}}

\change{We next evaluate the performance gains that can be achieved by opportunistic communications in the four baseline scenarios identified earlier and validate the optimal strategy to confirm that it achieves the highest possible gains.} Fig.~\ref{fig:bar_gain} gives the performance obtained for the four baseline scenarios with: \begin{inparaenum}[(i)] \item the optimal $d$ value provided by Theorem~\ref{thm:opt_number}, labeled \emph{Optimal Strategy}, \item the strategy proposed in Section~\ref{subsec:adapt_algo_basics} for the design of the adaptive algorithm, labeled $d = s$, and \item the adaptive algorithm implemented by HYPE, labeled \emph{HYPE}\end{inparaenum}. For each strategy, the figure shows the absolute average load of the cellular network in number of chunk copies per round ($D$).

\change{The results obtained show that the proposed approach can reduce very substantially the load of the cellular network (with offloaded traffic ranging from almost $50\%$ in the social data scenario to more than $95\%$ in the road traffic one).} The tests also show that the adaptive algorithm implemented by HYPE is very effective in minimizing this load, as it performs practically as the benchmarks given by the optimal and $d = s$ strategies.




\subsection{Impact of heterogeneity and sparsity}

\begin{figure}[t]
\centering
\includegraphics[width=0.49\textwidth]{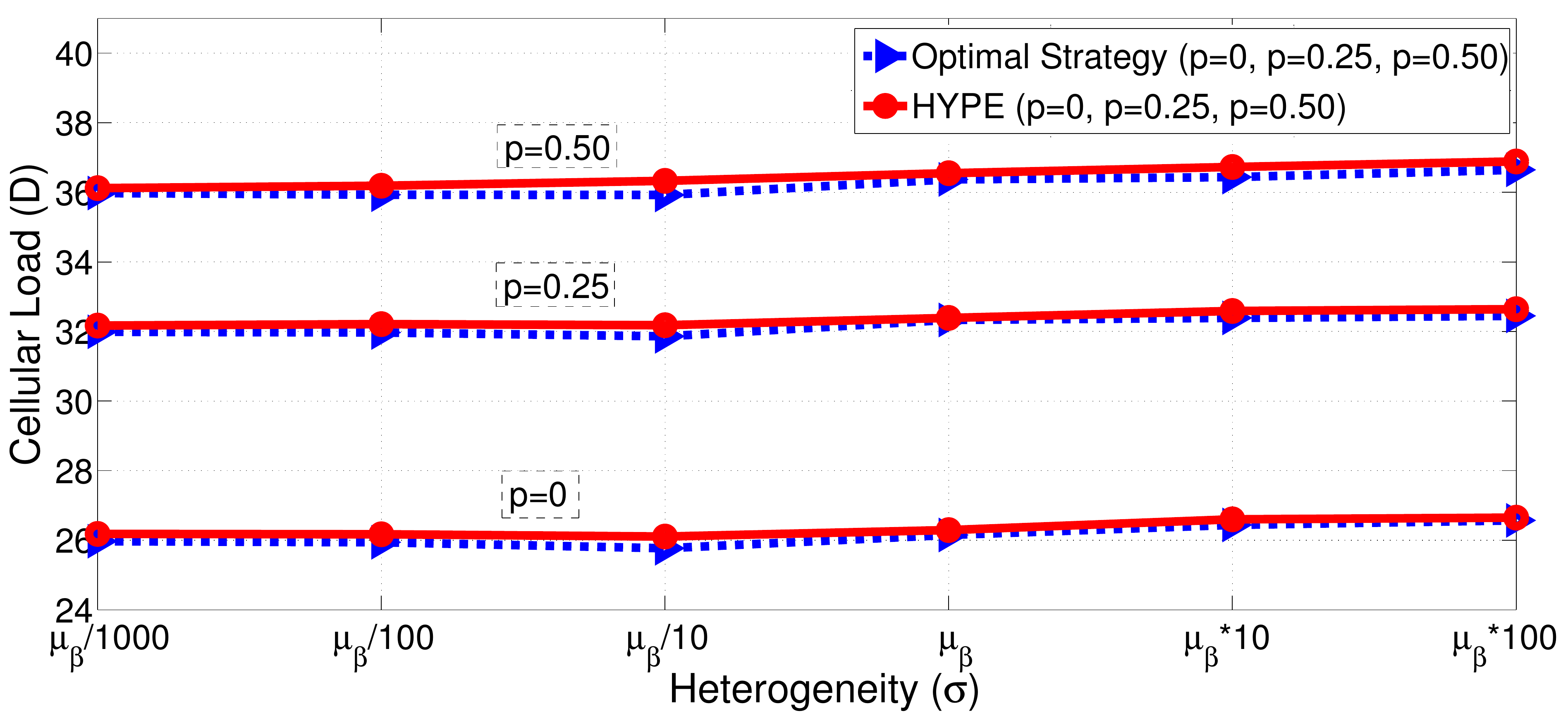}
 \caption{Cellular load $D$ as a function of the level of heterogeneity ($\sigma$) and network sparsity ($p$). }
  \label{fig:heterogeneous_plot}
\end{figure}

To understand the impact of the heterogeneity of pairwise contact rates $\beta_{xy}$ on the proposed approach, Fig.~\ref{fig:heterogeneous_plot} depicts the total cellular load $D$ for the streaming scenario, with varying $\sigma$'s. The effect of network sparsity is also shown by using different values for the the probability $p$ that a pair of nodes never meet.

We note that HYPE achieves a performance very close to the optimal, which confirms that the HYPE design also works for heterogeneous settings, as well as sparse ones. In the sparsest tested scenario ($p=0.50$), $D$ increases by $\approx38\%$ as compared to $p=0$, as a result of the slower dissemination caused by the decreasing number of connections (i.e., larger $p$). Furthermore, for all tested $p$ values, the cellular load $D$ is mostly insensitive to variations of $\sigma$ both for HYPE and the optimal strategy, which is in line with Theorem 1.

\subsection{Stability and response time}\label{sec-stability}

\begin{figure}[t]
\centering
 \includegraphics[width=0.49\textwidth]{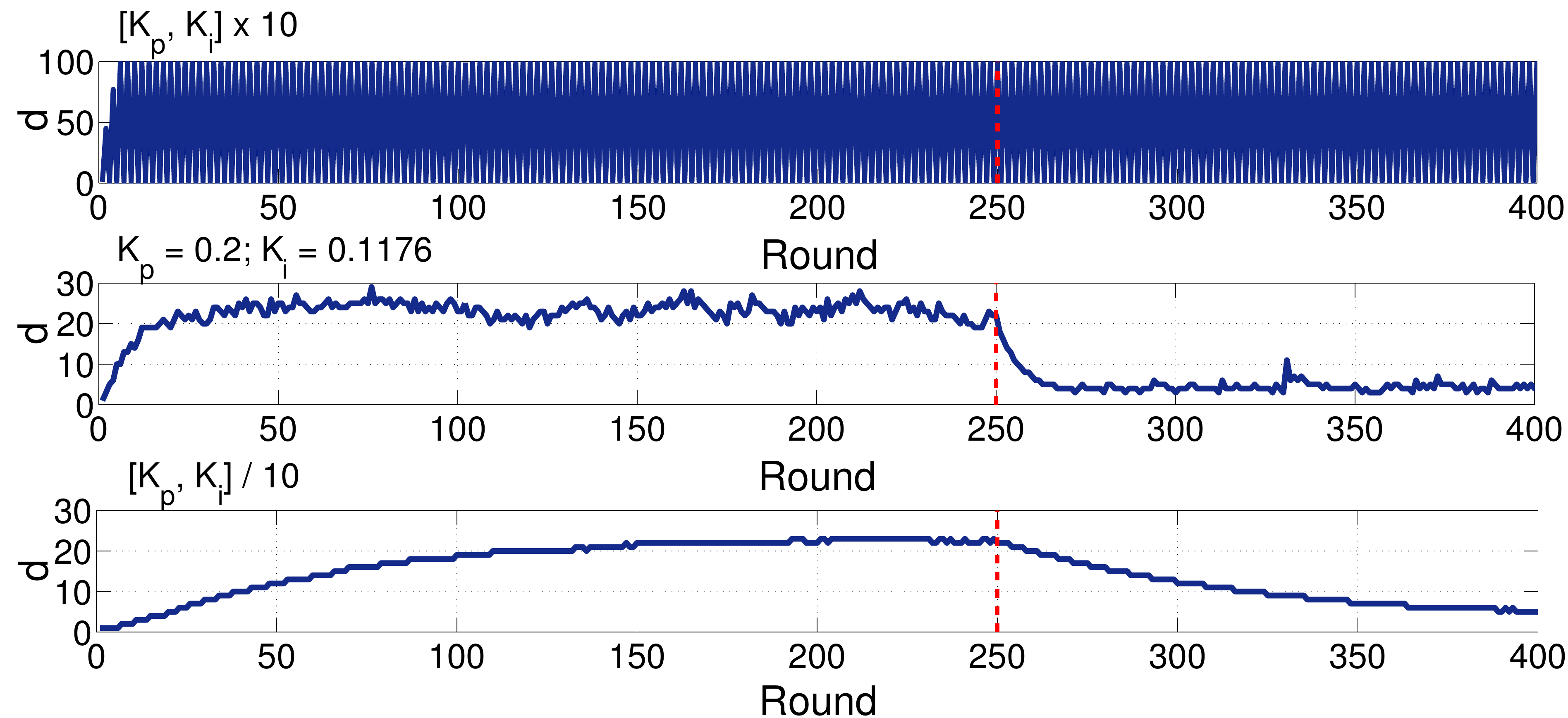}
 \caption{Evolution of the control signal $d$ over time for different $K_p,K_i$ settings. Our selection of parameters is stable and reacts quickly.}
  \label{fig:response}
\end{figure}

Based on the control theoretic analysis conducted in Section~\ref{sec:opt_str}, the parameters $K_p,K_i$ of the PI controller have been chosen to guarantee stability and ensure a good response time. In order to assess the effectiveness of this configuration, we evaluate its performance for the streaming baseline scenario and compare it against different choices for the values of parameters $K_p,K_i$. In Fig.~\ref{fig:response}, we show the evolution of the control signal $d$ over time for our setting $K_p = 0.2, K_i = 0.1176$, as well as a setting of these parameters ten times larger, labeled $[K_p, K_i]\times10$ and ten times smaller, labeled $[K_p, K_i]/10$. 
In the test, $\mu_\beta$ increases from $13$ contacts/pair/day to $40$ contacts/pair/day after $250$ rounds. (For instance, this could be the result of an increase in the number of contacts at rush hour).
Results show that our setting is stable and reacts quickly, while a larger setting of $K_p,K_i$ is highly unstable and a smaller setting reacts very slowly. This confirms the choice of parameters made for our controller. \add{We also conducted a similar experiment in which we varied $N$ (which could be for instance the result of a change in content popularity) and observed a similar behavior (not shown in the figure for space reasons).}


\begin{figure}[!t]
\centering
		\includegraphics[width=0.49\textwidth]{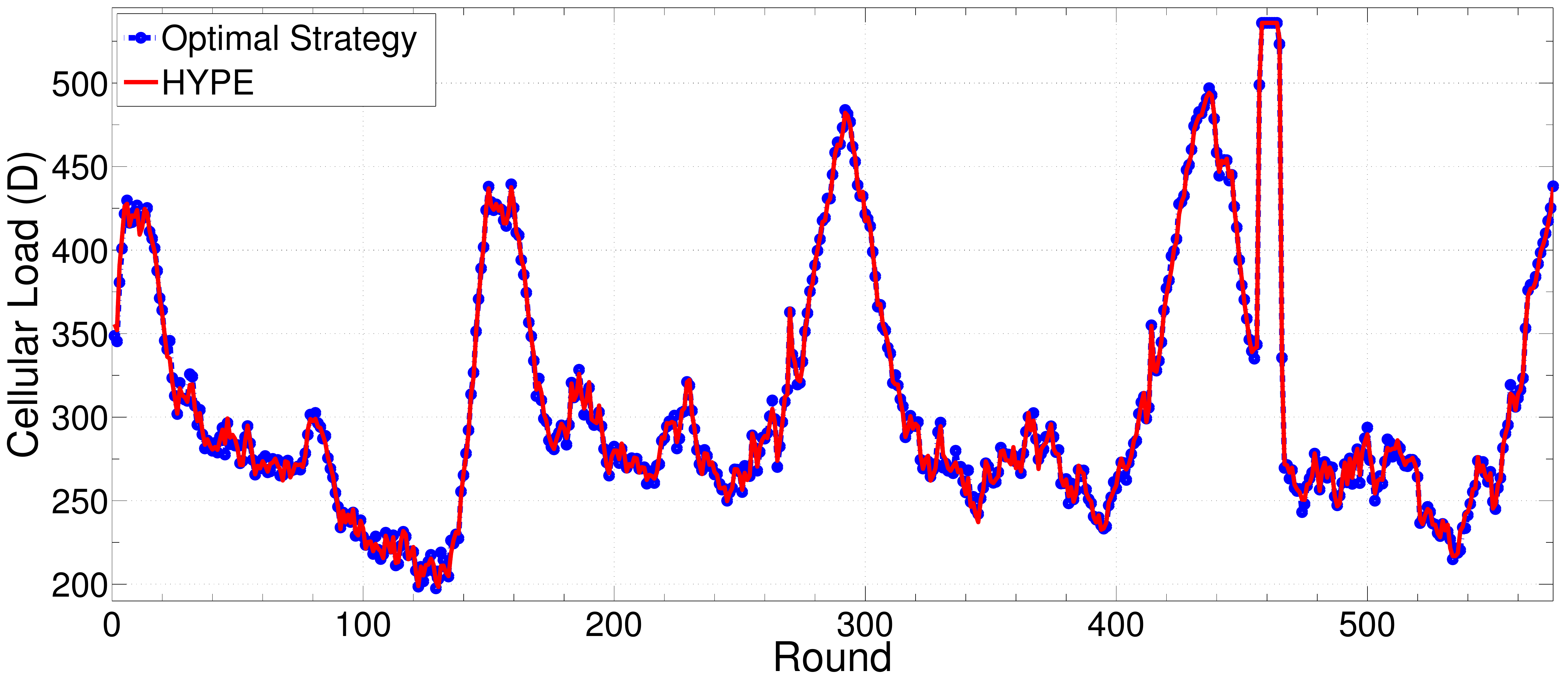}
		\caption{San Francisco real traces ($N=536$): Temporal evolution of the cellular load $D$ for HYPE and optimal strategy, with deadline $T_c=600$\,sec.}
		\label{fig:alldays_hype_vs_dinoptimal_D}
\end{figure}

\add{The above experiment shows the response of the controller to a drastic change.} In order to confirm that this response is sufficiently quick to follow the variations of the opportunistic contacts in a realistic environment, we consider the San Francisco real traces and study the temporal evolution of the cellular load $D$. To provide a benchmark, we compare HYPE against an optimal strategy that selects the best $d$ value every ten rounds.\footnote{For the optimal strategy, we make an exhaustive search over all possible $d$ values every ten rounds and select the best one. Note that such a strategy cannot be used in practice and is only considered for comparison purposes.} The results, for a content deadline $T_c=600$\,s, are plotted in Fig.~\ref{fig:alldays_hype_vs_dinoptimal_D}. These results confirm that HYPE reacts rapidly to dynamic conditions: as there are fewer number of contacts during night time, HYPE needs to inject more content through the cellular network (up to $D \approx  N$), while the higher number of contacts during day time greatly reduces the network load ($D\approx 220$).

\subsection{When to deliver: HYPE strategy versus other approaches}

\begin{figure}[t]
\centering
 \includegraphics[width=0.49\textwidth]{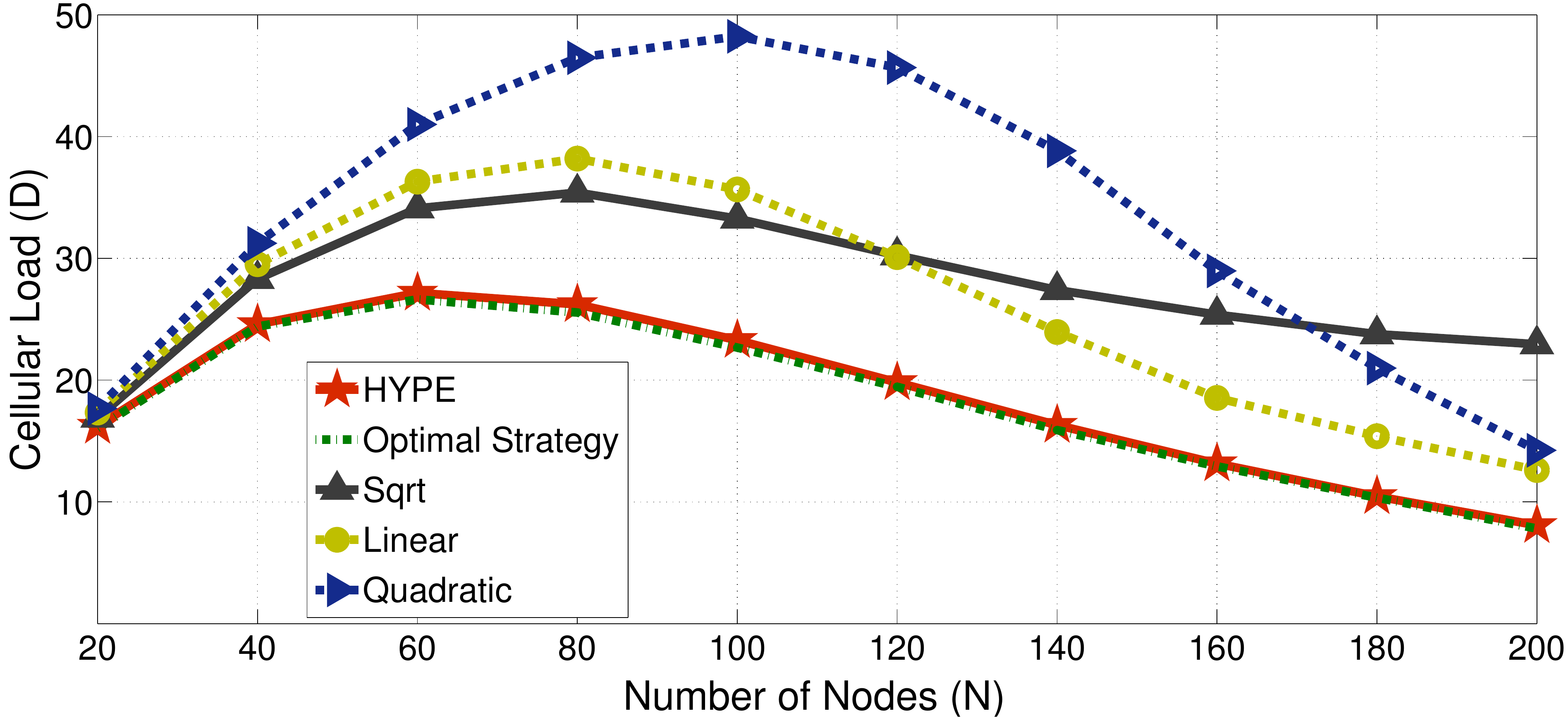}
 \caption{Comparison with Push-and-Track heuristics~\cite{Whitbeck2011}.}
  \label{fig:feedback}
\end{figure}

One of our key findings in Section~\ref{sec:opt_str} is that performance is optimized when all the deliveries over the cellular network take place at the beginning and at the end of the period. To validate this result, Fig.~\ref{fig:feedback} compares the performance of HYPE against the Push-and-Track heuristics proposed in~\cite{Whitbeck2011} (namely, \emph{Sqrt}, \emph{Linear} and \emph{Quadratic}), which distribute the deliveries along the period. Results are given for the social data scenario with a varying number of subscribed users $N$.
We observe from the figure that HYPE substantially outperforms all other approaches (the cellular load is even halved, in some cases), and performs very closely to the \emph{Optimal Strategy} benchmark.

\begin{figure}[t]
\vspace{-3mm}
\centering
\subfigure[San Francisco real traces ($N=536$)]{\includegraphics[width=0.49\textwidth]{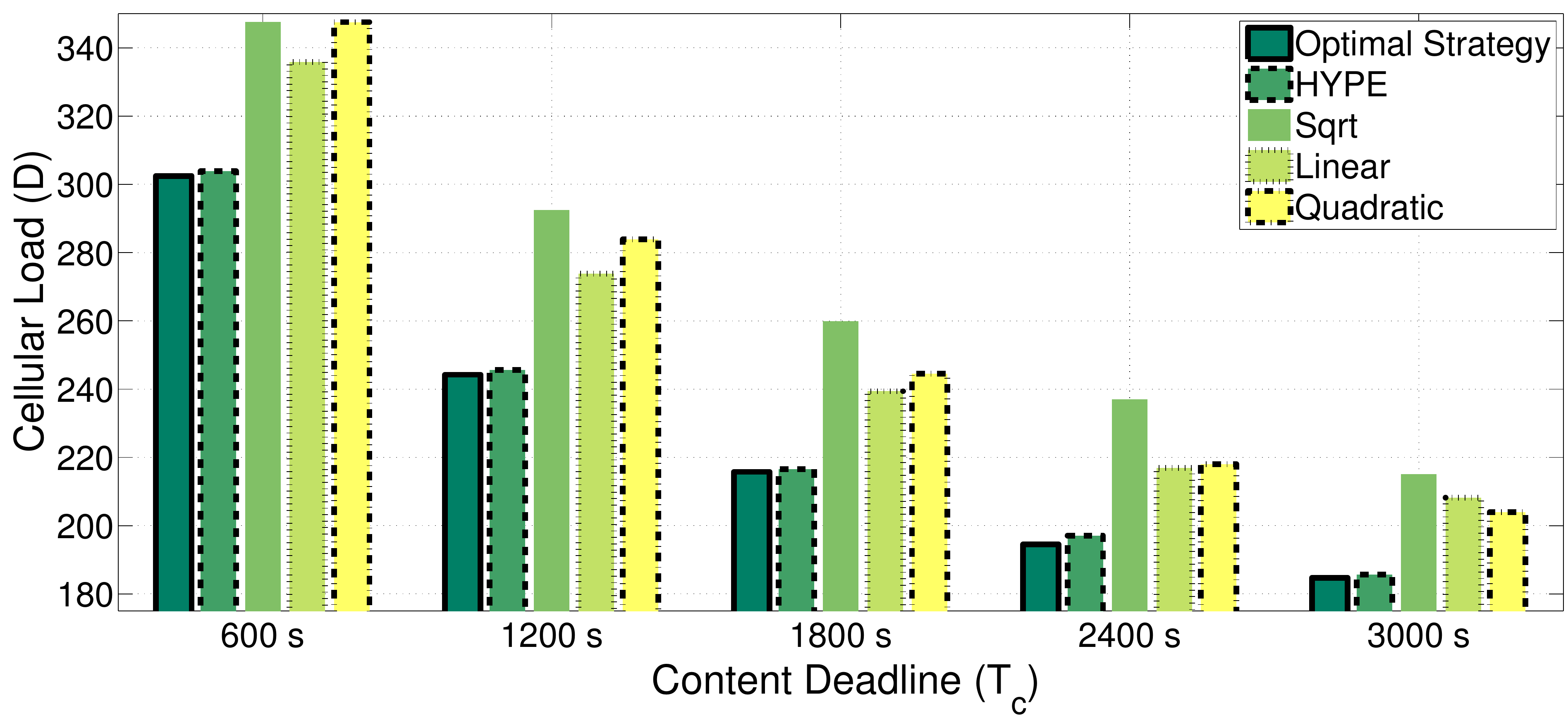}
\label{fig:sanfrancisco_traces}}
\subfigure[Infocom 2006 real traces ($N=78$)]{\includegraphics[width=0.49\textwidth]{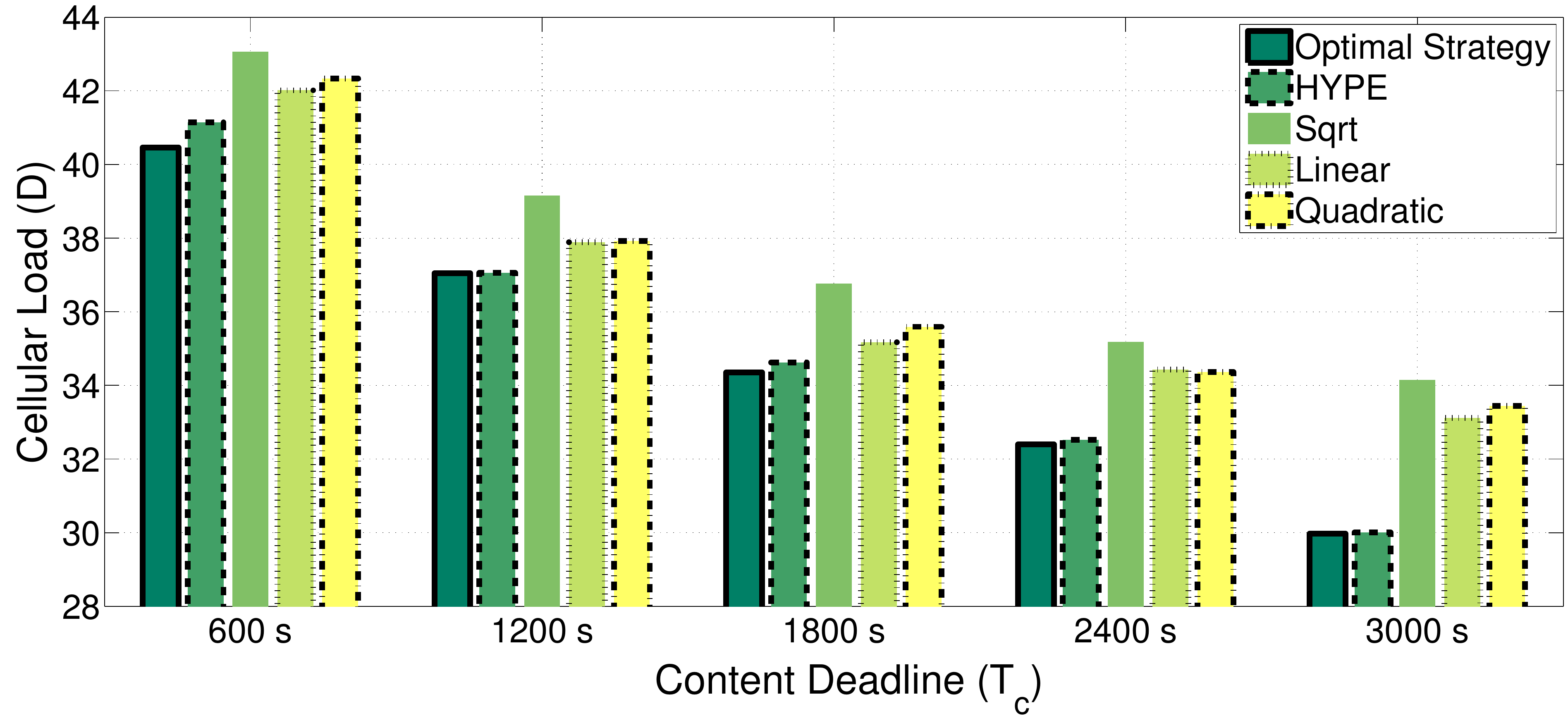}
\label{fig:infocom_traces}}
 \caption{Tests using real mobility traces for different deadlines $T_c$. HYPE performs closely to the benchmark provided by the optimal strategy and substantially outperforms previous heuristics.}
 \label{fig:fig:real_trace_benchmark}
\end{figure}

In addition to the above experiment, conducted with a mobility model, we also compare the performance of HYPE against the other approaches with real mobility traces. The results, depicted in Fig.~\ref{fig:fig:real_trace_benchmark}, show that HYPE closely follows the performance of the benchmark given by the optimal strategy and outperforms previous heuristics. For the San Francisco real traces, HYPE can offload about $20\%$ more traffic than the previous heuristics. For the Infocom 2006 traces, the employed strategy has a smaller impact on cellular load performance, which yields to a smaller gain (up to about $12$\%).\footnote{\add{Indeed, by conducting experiments with the Infocom 2006 traces for many different strategies (unreported here for space reasons), we observed that performance was relatively similar for all of them, which shows that the impact of the specific strategy followed is limited for this case.}}


\subsection{Which seed nodes: comparison to other selection methods}

One of the key decisions in the HYPE design is to randomly select a node when transmitting content over the cellular network. In order to gain insight into the impact of this design decision, we compare HYPE against the heuristic approach proposed in~\cite{Han2010} to select the seed nodes in the opportunistic network. Unlike HYPE,~\cite{Han2010} requires full knowledge of the pairwise contact rates to identify the target set of users, which involves a much higher level of complexity. Note that, since~\cite{Han2010} does not provide an algorithm to compute the number of copies $d$ of the data chunk, we apply the HYPE strategy to compute $d$ also for~\cite{Han2010}.

\begin{figure}[t]
\centering
 \includegraphics[width=0.49\textwidth]{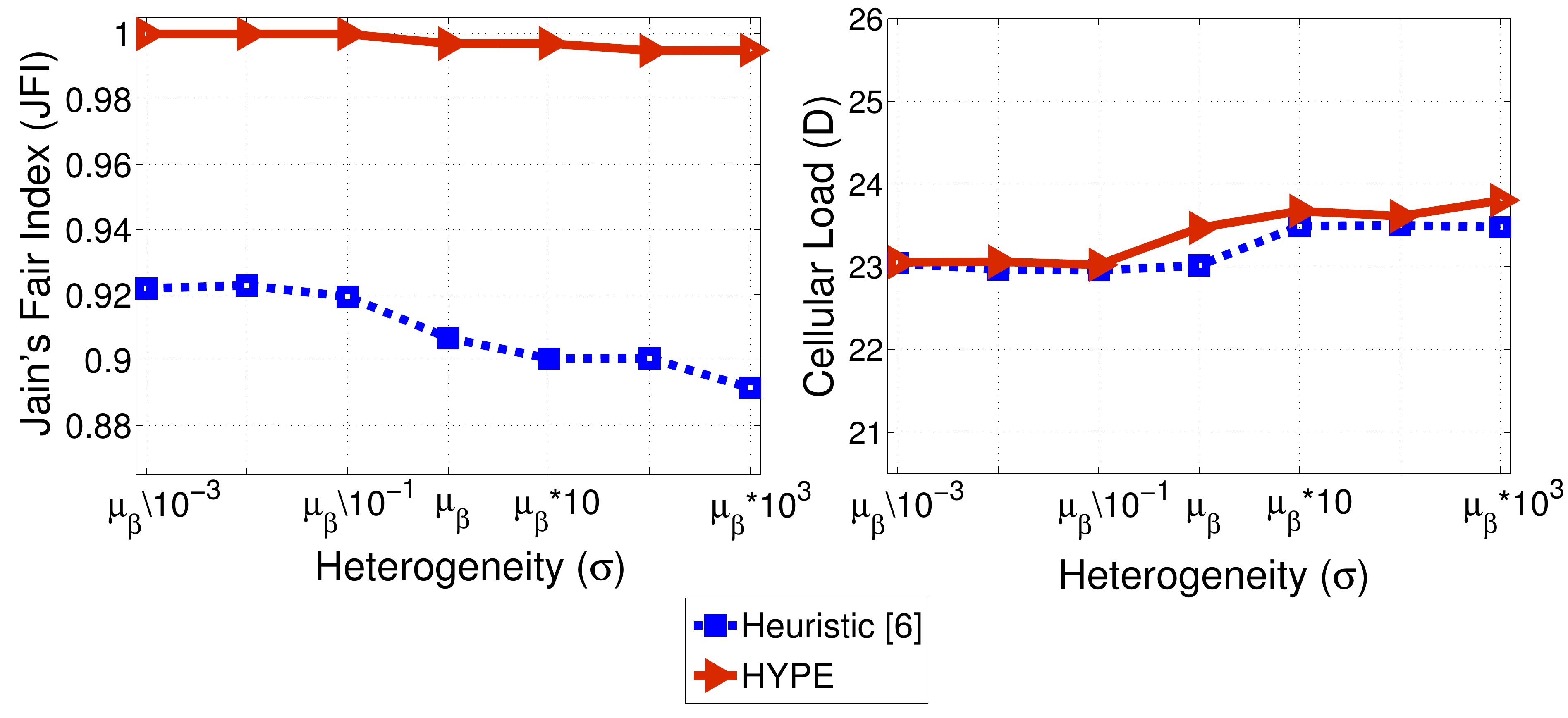}
 \caption{HYPE versus the heuristic solution of~\cite{Han2010}, varying the heterogeneity ($\sigma$). HYPE provides a much better trade-off between fairness and cellular load performance.}%
  \label{fig:selection_algorithm}
\end{figure}

Fig.~\ref{fig:selection_algorithm} shows the performance of both approaches in terms of cellular traffic load ($D$) and fairness for different values of heterogeneity $\sigma$, for the social data scenario. To measure fairness, we apply the Jain's Fairness Index ($JFI$) to the total number of cellular and opportunistic communications involving a node.\footnote{For instance, a node that \begin{inparaenum}[(i)]\item receives the content through the cellular network and \item sends it to $n$ nodes in range during the period, will have a total number of communications equal to $n+1$\end{inparaenum}.} The results show that HYPE provides a much higher level of fairness than~\cite{Han2010} with negligible loss in terms of cellular traffic load. Therefore, HYPE does not only feature a simpler implementation than~\cite{Han2010}, as it does not need to know the individual inter-contact rates, but also provides a much better trade-off between fairness and cellular load performance. The results of this and the previous section are particularly relevant as the algorithms of \cite{Whitbeck2011} and \cite{Han2010} are the only existing proposals in the literature to offload cellular networks with opportunistic communications while providing deterministic delay guarantees.



\subsection{Signaling load}
\begin{figure}[t]
\centering
\includegraphics[width=0.49\textwidth]{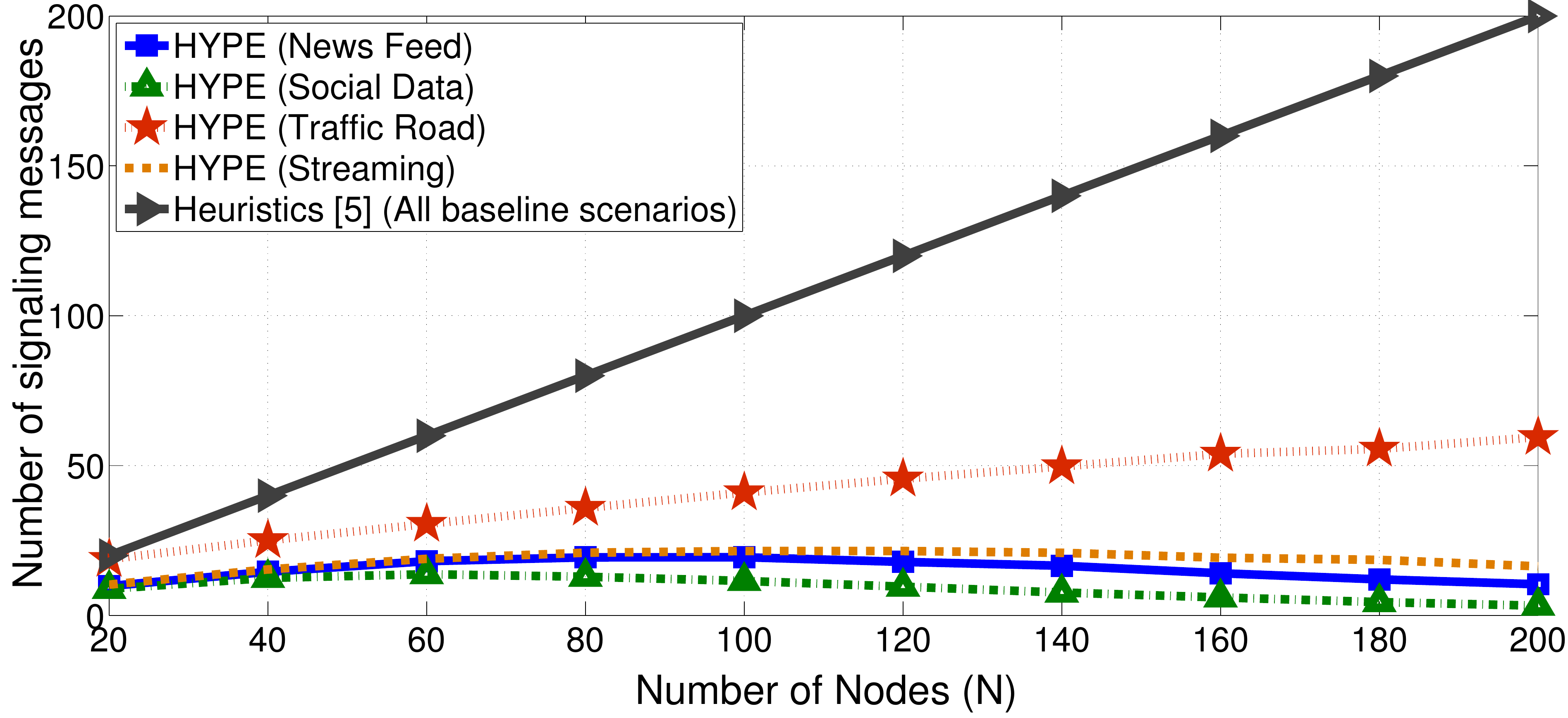}
 \caption{HYPE signaling load as a function of the number of nodes $N$ for the four baseline scenarios. HYPE outperforms the previous heuristics of ~\cite{Whitbeck2011} and scales very efficiently with the number of nodes.}
  \label{fig:signaling_load}
\end{figure}

In order to gain insight into the scalability of our design, we analyze the number of uplink signaling messages sent over cellular network. In HYPE, such messages are the signals sent by the nodes with a single copy \emph{ID} to the Content Server at the end of the period. Fig.~\ref{fig:signaling_load} shows the signaling load of HYPE as a function of the number of users for each of the four baseline scenarios, and compares it with the Push-and-Track heuristics (labeled as \emph{Heuristics~\cite{Whitbeck2011}} in the figure), which require an uplink signal each time a node receives the chunk.\footnote{We have not compared the signaling messages of HYPE against the approach of \cite{Han2010} since that approach requires gathering data from nodes' mobility patterns. Even though~\cite{Han2010} does not explain the signaling mechanism employed, we expect that the need to collect the mobility patterns involves a substantially higher signaling overhead than~\cite{Whitbeck2011}.} In contrast, HYPE scales very efficiently with the number of users: the more users are subscribed to the content, the lower the signaling load per user.\newline

Summarizing \add{the results of the performance evaluation conducted in} this section, we have shown that our analytical model is very accurate, that the optimal strategy proposed does indeed minimize the load incurred in the cellular network, that the adaptive algorithm is stable and quick to converge to optimality and that HYPE outperforms existing heuristics in crucial aspects: cellular load, signaling load and fairness among users.

\section{Conclusion}\label{sec:conclusion}

In this paper we have presented HYPE, a novel approach to offload cellular traffic through opportunistic communications.
To design HYPE, we have developed a theoretical model to analyze the performance of opportunistic dissemination when data
can be selectively injected through a cellular network. Based on this model, we have
derived the optimal strategy that minimizes the total amount of data injected through the cellular network while meeting delay guarantees.
To implement the optimal strategy obtained from the analysis, HYPE runs an adaptive algorithm
that adjusts the data delivery over the cellular network to the current network conditions. By building on control
theory, we guarantee that this algorithm is stable and quickly adapts to dynamic conditions. The algorithm incurs
very low signaling overhead and does not need to monitor the contacts between nodes nor to gather complex statistics, which are important
requirements for a practical implementation.

\section*{Acknowledgments}

\change{We thank the anonymous reviewers for their inspiring and insightful feedbacks.} The project has been sponsored by the Microsoft Innovation Cluster for Embedded Software (ICES) and by the CROWD project of the EU Seventh Framework Programme (FP7).

\bibliographystyle{IEEEtran}

\small
\bibliography{hycloud}

\normalsize
\vspace{-5mm}
\appendix
\label{appendix}

\setcounter{theorem}{0}
\setcounter{proposition}{0}

\begin{theorem}\label{thm:mc_diff_eq_app}
According to the HYPE Markov chain for heterogeneous mobility (similar to Fig.~\ref{fig:mc_hetero}), the process $\{M(t),t \ge 0\}$ is described by the following system of differential equations:
\begin{equation}
\begin{cases}
\frac{d}{d t} p_1^{c_1}(t) = - \lambda_1 p_1^{c_1}(t),& i= 1\\
\frac{d}{d t} p_i^{c_1}(t) = - \lambda_{i} p_{i}^{c_1}(t) + \lambda_{i-1} p_{i-1}^{c_1}(t), &  1 < i < N\\
\frac{d}{d t} p_N^{c_1}(t) = \lambda_{N-1} p_{N-1}^{c_1}(t),&  i= N
\end{cases}
\label{equ:birth_diff_equ_app}
\end{equation}
where $\lambda_{i} = i (N-i) \mu_{\beta}$. (Recall that $\mu_{\beta}$ is the known expectation of the generic probability distribution $F(\beta):(0,\infty)\rightarrow[0,1]$, from which the inter-contact rates describing our network are drawn: $\{\beta_{xy}\} = \mathbf{B}$.)
\end{theorem}
\begin{IEEEproof}
Recall that we denoted by $\{\mathbf{K}^{i}_{1},\mathbf{K}^{i}_{2},\ldots,\mathbf{K}^{i}_{{N\choose i}}\}$ the set of ${N\choose i}$ states in the Markov chain corresponding to level $i$. Also, $\mathbf{B}= \{\beta_{xy}\}$ is our network. Then, assuming the Markov chain starts in initial state $\mathbf{K}_{m}^{1}$ for $1 \leqslant m\leqslant N$ (i.e., the chunk was initially injected to a node $m$), the probability of still being in this state after a small time interval $dt$ is:
\begin{equation}\label{eq:mc_diff_1}
\mathbb{P}\left[\mathbf{K}_{m}^{1} \text{ at } t+\mathrm{d}t\;|\;\mathbf{B}\right] = \mathbb{P}\left[\mathbf{K}_{m}^{1} \text{ at } t\;|\;\mathbf{B}\right]\cdot\left(1-\sum_{\substack{x\in \mathbf{K}_{m}^{1}\\y\notin \mathbf{K}_{m}^{1}}} \beta_{xy}\,\mathrm{d}t\right)
\end{equation}

Then, averaging over all states of this dissemination level, the probability of still being at dissemination level $1$ after a small time interval $dt$ is:
\begin{equation}
\mathbb{P}\left[1 \text{ at } t+\mathrm{d}t\;|\;\mathbf{B}\right] = \sum_{m=1}^{N} \mathbb{P}\left[\mathbf{K}_{m}^{1} \text{ at } t+\mathrm{d}t\;|\;\mathbf{B}\right]
\end{equation}

Finally, considering all possible network realizations:
\begin{align}
p_1^{c_1}(t+dt) &= \mathbb{P}\left[1 \text{ at } t+\mathrm{d}t\right] \nonumber\\
& = \int_{\mathbf{B}} \sum_{m=1}^{N} \mathbb{P}\left[\mathbf{K}_{m}^{1} \text{ at } t+\mathrm{d}t\;|\;\mathbf{B}\right] \mathbb{P}\left[\mathbf{B}\right]\mathrm{d}\mathbf{B},
\end{align}
where $\mathbb{P}\left[\mathbf{B}\right]$ is given by the generic distribution, $F(\beta):(0,\infty)\rightarrow[0,1]$, which determines the inter-contact rates of our network. Combining this last equation with Eq.~(\ref{eq:mc_diff_1}) and using basic probability theory, we obtain Eq.~(\ref{eq:mc_diff_2}) below:
\begin{align}
& p_1^{c_1}(t+dt) \nonumber\\=&\; \int_{\mathbf{B}} \sum_{m=1}^{N} \mathbb{P}\left[\mathbf{K}_{m}^{1} \text{ at } t\;|\;\mathbf{B}\right]\left(1-\sum_{\substack{x\in \mathbf{K}_{m}^{1}\\y\notin \mathbf{K}_{m}^{1}}} \beta_{xy}\,\mathrm{d}t\right)\mathbb{P}\left[\mathbf{B}\right]\mathrm{d}\mathbf{B} \nonumber\\
=&\;\;\int_{\mathbf{B}} \sum_{m=1}^{N} \mathbb{P}\left[\mathbf{K}_{m}^{1} \text{ at } t\;|\;\mathbf{B}\right]\mathbb{P}\left[\mathbf{B}\right]\mathrm{d}\mathbf{B}\;- \nonumber\\
&- \int_{\mathbf{B}} \sum_{m=1}^{N} \mathbb{P}\left[\mathbf{K}_{m}^{1} \text{ at } t\;|\;\mathbf{B}\right]\sum_{\substack{x\in \mathbf{K}_{m}^{1}\\y\notin \mathbf{K}_{m}^{1}}} \beta_{xy} \,\mathrm{d}t\cdot\mathbb{P}\left[\mathbf{B}\right]\,\mathrm{d}\mathbf{B} =  p_1^{c_1}(t) - \nonumber\\
&\;\;\sum_{m=1}^{N} \int_{\mathbf{B}}\sum_{\substack{x\in \mathbf{K}_{m}^{1}\\y\notin \mathbf{K}_{m}^{1}}} \beta_{xy}\,\mathrm{d}t \cdot \mathbb{P}\left[\mathbf{B}\;|\;\mathbf{K}_{m}^{1} \text{ at } t\right] \mathbb{P}\left[\mathbf{K}_{m}^{1} \text{ at } t\right] \,\mathrm{d}\mathbf{B} \nonumber\\
=&\;\;p_1^{c_1}(t) - \sum_{m=1}^{N} \mathbb{P}\left[\mathbf{K}_{m}^{1} \text{ at } t\right] \mathbb{E}\left[X\;|\;\mathbf{K}_{m}^{1} \text{ at } t\right]\,\mathrm{d}t,\label{eq:mc_diff_2}
\end{align}
where $X = \sum_{\substack{x\in \mathbf{K}_{m}^{1}\\y\notin \mathbf{K}_{m}^{1}}} \beta_{xy}$ (that is a sum of $N-1$ terms).

Since our network's inter-contact rates forming the matrix $\mathbf{B}$ are independent and identically distributed (with generic distribution $F(\beta):(0,\infty)\rightarrow[0,1]$ of mean $\mu_{\beta}$), the terms of the sum forming $X$ are distributed according to $F(\beta)$, regardless of the specific node combination $\mathbf{K}_{m}^{1}$. Hence, $\mathbb{E}\left[X\;|\;\mathbf{K}_{m}^{1} \text{ at } t\right] = \mathbb{E}\left[X\right]$ and Eq.~(\ref{eq:mc_diff_2}) becomes:
\begin{align}
p_1^{c_1}(t+dt) &= p_1^{c_1}(t) - \mathbb{E}\left[X\right]\,\mathrm{d}t\cdot\sum_{m=1}^{N} \mathbb{P}\left[\mathbf{K}_{m}^{1} \text{ at } t\right]\\
&= p_1^{c_1}(t) - (N-1)\mu_{\beta}\,\mathrm{d}t\cdot p_1^{c_1}(t).
\end{align}

Thus, we obtain as desired:
\begin{equation}
\frac{d}{d t} p_1^{c_1}(t) = - (N-1)\mu_{\beta} p_1^{c_1}(t)
\end{equation}

The remaining two differential equations are obtained by the same process.
\end{IEEEproof}

\begin{theorem}In the optimal strategy, the data chunk is delivered through the cellular network to $d$ seed nodes at time $t = 0$, and to the nodes that do not have the content by the deadline $t=T_c$.
\end{theorem}
\begin{IEEEproof}The proof goes by contradiction: we first assume that in the optimal strategy the data chunk is transmitted
 to some mobile node at time $t \neq \{0, T_c\}$ and then we find an alternative strategy that provides a better
performance.

If the chunk is transmitted to some mobile node at $t \neq \{0, T_c\}$, this means that $C \neq \{1,\ldots,d\}$ and hence 
there exists some missing value smaller than $c_d$ in $C$. Indeed, if $C = \{1,\ldots,d\}$, all the first $d$ states are
instantaneous states and the data chunk is transmitted to $d$ nodes at the beginning of the round.

Let us denote the largest value in $C$ ($c_d$) by $k$ and the largest value that is missing by $k-l$. Let $D_{k}$ further 
denote the value of $D$ for the optimal configuration $C_k = \{c_1,\ldots,c_{d}\}$ (where $c_d = k$), $D_{k-l}$ the
value of $D$ for the configuration $C_{k-l} = \{c_1,\ldots,c_{d-1},k-l\}$ and $D_{k+1}$ the value of $D$ for the
configuration $C_{k+1} = \{c_1,\ldots,c_{d-1},k+1\}$.\footnote{Without loss of generality we assume that $c_d \neq N$,
as it can be easily shown that a configuration with $c_d = N$ is not optimal.} In the following, we show that either
$D_{k-l}$ or $D_{k+1}$, or both, are smaller than $D_{k}$, which contradicts the initial assumption that the
configuration $\{c_1,\ldots,c_d\}$ is optimal.

If we compare the state probabilities for the configurations $C_k$ and $C_{k+1}$, we have that
\begin{equation}
\begin{cases}
P_{i}^{C_k}(s) = P_{i}^{C_{k+1}}(s), & i < k\\
P_{i}^{C_k}(s) = \displaystyle\frac{\lambda_{k+1} (s + \lambda_{k})}{\lambda_{k} (s + \lambda_{k+1})} P_{i}^{C_{k+1}}(s), & i > k+1.
\end{cases}
\end{equation}

From the above, we have that the following holds for $i > k+1$, 
\begin{align}
P_{i}^{C_k}(s) - P_{i}^{C_{k+1}}(s) =&\;\left(\frac{\lambda_{k+1}}{s+\lambda_{k+1}} -
\frac{\lambda_{k}}{s+\lambda_{k}}\right) \prod_{j \in S_{i-1}^{C_k}\setminus k}{\frac{\lambda_j}{s+\lambda_j}}\nonumber\\
=&\;\frac{\lambda_{k+1} - \lambda_{k}}{\lambda_{k+1}\lambda_{k}} s P_i^{C_{d-1}}(s),
\end{align}
where $ S_{i-1}^{C_k}=\{1,2,\ldots,i-1\} \setminus (\{1,2,\ldots,i-1\} \cap \{c_1,\ldots,c_d\})$ and
$P_i^{C_{d-1}}(s)$  is state $i$'s probability for the configuration $C_{d-1} = \{c_1,\ldots,c_{d-1}\}$.

By doing the inverse Laplace transform of the above, we have that
\begin{align}\label{eq-16}
p_{i}^{C_k}(t) - p_{i}^{C_{k+1}}(t) = \frac{\lambda_{k+1} - \lambda_{k}}{\lambda_{k}\lambda_{k+1}} \frac{d P_i^{C_{d-1}}(t)}{d t} 
= &\;\frac{\lambda_{k+1} - \lambda_{k}}{\lambda_{k}\lambda_{k+1}} \cdot \nonumber\\ \left(- \lambda_{i}P_i^{C_{d-1}}(t) + \lambda_{i-1}P_{i-1}^{C_{d-1}}(t)\right).
\end{align}

Furthermore, we also have
\begin{equation}
P_{k+1}^{C_k}(s) - P_{k}^{C_{k+1}}(s) = \frac{\lambda_{k} - \lambda_{k+1}}{\lambda_{k}} P_{k+1}^{C_{d-1}}(s),
\end{equation}
and, hence,
\begin{equation}\label{eq-17}
p_{k+1}^{C_k}(t) - p_{k}^{C_{k+1}}(t) = \frac{\lambda_{k} - \lambda_{k+1}}{\lambda_{k}} p_{k+1}^{C_{d-1}}(t).
\end{equation}

Combining Eqs.~(\ref{eq-16}) and (\ref{eq-17}) with Eq.~(\ref{eq-D}), and taking into account that for $i > k+1$ it
holds that $d_i + d^*_i = d_{i+1} + d^*_{i+1} +1$, we obtain
\begin{align}
D_{k} - D_{k+1} =  (\lambda_{k} - \lambda_{k+1}) \sum_{i =
k+1}^{N-1}{{\frac{1}{\lambda_{k}\lambda_{k+1}}}p_i^{C_{d-1}}(T_c)}.
\end{align}

Following a similar approach for the configurations $C_k$ and $C_{k-l}$, we obtain
\begin{align}
D_{k} - D_{k-l} =  (\lambda_{k} - \lambda_{k-l}) \sum_{i =
k+1}^{N-1}{{\frac{1}{\lambda_{k}\lambda_{k-l}}}p_i^{C_{d-1}}(T_c)}.
\end{align}

Since it holds that either $\lambda_{k} - \lambda_{k-l}$ or $\lambda_{k} - \lambda_{k+1}$ is greater than zero, we have
that at least one of the two alternative configurations ($C_{k+1}$ or $C_{k-l}$) provides a $D$ value smaller than $C_{k}$. This 
contradicts the assumption that in the optimal strategy the data chunk is transmitted to some node at
time $t \neq \{0, T_c\}$, which proves the theorem.
\end{IEEEproof}

\vspace{0.3cm}

\begin{proposition} Let us define $G_d$ as the gain resulting from sending the $(d+1)^{th}$ chunk of chunk copy at the beginning of the period (i.e., $G_d = D_{d} - D_{d+1}$, where $D_{d+1}$ and $D_{d}$ are the values of $D$ when we deliver $d+1$ and $d$ copies at the beginning, respectively). Then, $G_d$ can be computed from the following equation:
\begin{equation}
G_d = \sum_{j = d}^{N-1}{\frac{\lambda_j}{\lambda_{d}}p_{j}^d(T_c)} - 1.
\end{equation}
\end{proposition}

\begin{IEEEproof}
$G_d$ can be expressed as:
\begin{equation}\label{eq-G}
G_d = \sum_{j = d}^{N-1}{(N-j)\left(p_{j}^d(T_c) - p_{j}^{d+1}(T_c)\right)} - 1.
\end{equation}

The term $p_{j}^d(T_c) - p_{j}^{d+1}(T_c)$ is calculated as follows. From Eqs.~(\ref{eq-laplace}), we have
\begin{equation}\label{eq-si+1j}
P_{j}^d(s) - P_{j}^{d+1}(s) = - \frac{s P_{j}^d(s)}{\lambda_{d}}.
\end{equation}

Making the inverse Laplace transform of the above for $j > d$ yields
\begin{align}\label{eq-20}
p_{j}^d(T_c) - p_{j}^{d+1}(T_c) & = -\frac{1}{\lambda_d} \frac{d p_{j}^d(t)}{d t}\bigg|_{T_c} = \\\nonumber
& = \frac{1}{\lambda_{d}}(\lambda_j p_{j}^d(T_c) - \lambda_{j-1} p_{j-1}^d(T_c)).
\end{align}

Note that the above equation also holds for $j = d$ since in this case $p_{j-1}^d(t) = 0$ and $p_{j}^{d+1}(t) = 0$. 
Combining it with Eq.~(\ref{eq-G}) leads to
\begin{equation}\label{eq-22}
G_d = \sum_{j = d}^{N-1}{\frac{\lambda_j}{\lambda_{d}}p_{j}^d(T_c)} - 1.
\end{equation}
\end{IEEEproof}

\vspace{0.3cm}

\begin{theorem}
The optimal value of $d$ is the one that satisfies $G_d  = 0$.
\end{theorem}

\begin{IEEEproof}
As long as $G_d > 0$, we benefit from increasing $d$, since by sending one additional chunk at the beginning, we save
more than one chunk at the end. Conversely, if $G_d < 0$ we do not benefit. It can be seen that $G_1 > 0$ and $G_N < 0$.
Furthermore, it can also be seen that $G_d$ strictly decreases with $d$:
\begin{align}
G_{d+1} - G_d =&\;\sum_{j = d}^{N-1}{\displaystyle\frac{\lambda_j}{\lambda_{d+1}} p^{d+1}_{j}(T_c) - \frac{\lambda_j}{\lambda_d}
p^{d}_{j}(T_c)} \\
=&\;\sum_{j = d}^{N-1}{\displaystyle\frac{p^{d}_{j}(T_c)\lambda_j(\lambda_{j+1} - \lambda_j - (\lambda_{d+1}-\lambda_{d}))}{\lambda_d
\lambda_{d+1}}} < 0.\nonumber
\end{align}

From the above, it follows that the value of $d$ that minimizes $D$ is the one that satisfies $G_d = 0$, since up to
this value we benefit from increasing $d$ and after this value we stop benefiting, which proves the theorem.
\end{IEEEproof}

\vspace{0.3cm}
\begin{theorem} The HYPE control system is stable for $K_p = 0.2$ and $K_i = 0.4/3.4$.
\end{theorem}
\begin{IEEEproof}
The closed-loop transfer function of our system is
\begin{eqnarray}\label{eq-tf}
T(z)  =  \frac{- z(z-1) H K_p - z H K_i}{z^2 + (- H K_p -1)z + H (K_p - K_i)}
\end{eqnarray}
where 
\begin{equation}\label{eq-h}
H = - 2
\end{equation}

A sufficient condition for stability is that the poles of the above polynomial fall within the unit circle $|z| < 1$. This can be ensured by choosing coefficients $\{a1, a2\}$ of the characteristic polynomial that belong to the stability triangle~\cite{astrom}:
\begin{equation}\label{eq-tringle}
\begin{cases}
a_2 < 1\\
a_1 < a_2 + 1\\
a_1 > -1 - a_2
\end{cases}
\end{equation}

In the transfer function of Eq. (\ref{eq-tf}), the coefficients of the characteristic polynomial are $a_1 = - H K_p -1$ and $a_2 = H (K_p - K_i)$. From Eqs. (\ref{eq-kp}) and (\ref{eq-h}), we have $H K_p = - 0.4$ and $H K_i = - 0.4/(0.85 \cdot 2)$, from which $a_1 = - 0.6$ and $a_2 = -0.16$. It can be easily seen that these $\{a1, a2\}$ values satisfy Eq.~(\ref{eq-tringle}), which proves the theorem.
\end{IEEEproof}

\end{document}